\documentclass[usenatbib]{mnras}
\usepackage{txfonts}
\usepackage{natbib}
\usepackage{epsfig,rotating,amssymb}
\newcommand{\DXDYCZ}[3]{\left( \frac{ \partial #1 }{ \partial #2 }
                         \right)_{#3}
                        }
\def \kmsec {\rm km~s$^{-1}$}
\def \cmthree {cm$^{-3}$}
\def \cmtwo {cm$^{-2}$}
\def \al {\rm et al.~}
\def \13CO {$^{13}$CO}
\def \C18O {C$^{18}$O}
\def \Tkin {$T_{\rm kin}$}
\def \Tex {$T_{\rm ex}$}
\def \Tmb {$T_{\rm mb}\,$}
\def \Tcol {$T_{\rm col}\,$}
\def \Trstar {$T_{\rm r}$$^{\rm *}$}
\def \Tastar {$T_{\rm a}$$^{\rm *}$}
\def \Tsys {$T_{\rm sys}$}
\def \Ncol {N$_{\rm col}$}
\def \NHtwo {$N_{\rm{H_2}}$}
\def \Msun {$M_{\odot}$}
\def \Lsun {$L_{\odot}\thinspace$}
\def \arcsec {$^{\prime\prime}$}
\def \quote {$^{\rm{\prime}}$}
\def \arcmin {$^\prime$}
\def \mum {$~\mu$m\,}
\def \bsp {$\!\!$}
\def \j#1#2{$J=#1-#2\,$}
\def \iso#1#2{$^{#1#2}$\bsp}
\def \nmb {$\eta_{\rm mb}$}
\def \nfss {$\eta_{\rm fss}$}
\def \abun {N(X)/N$_{H2}$}
\def \MN {MNRAS}
\def \3P1 {$^3$P$_1$ -- $^3$P$_0$}
\def \Halpha {H$\alpha$}
\def \H2 {H$_{2}$}
\def \nh2 {n$_{H_2}$\,}
\def \Brg {Br$\gamma\,$}
\def \degree {$\ensuremath{^\circ}\,$}

\begin{document}
\label{firstpage}
\pagerange{\pageref{firstpage}--\pageref{lastpage}}
\title{A GMRT 610 MHz radio survey of the North Ecliptic Pole (NEP, ADF-N) / Euclid Deep Field North}
\author[Glenn J. White \al]
{Glenn J. White$^{1,2}$, L. Barrufet$^{1,2,3,4}$,  S. Serjeant$^1$, C.P. Pearson$^{2,1}$, C. Sedgwick$^1$, S. Pal$^{5,6,7}$,
\newauthor T.W. Shimwell$^{8,9}$, S.K. Sirothia$^{5,6}$, P. Chiu$^{2}$, N. Oi$^{10}$, T. Takagi$^{11}$, H. Shim$^{12}$, 
\newauthor H. Matsuhara$^{13}$, D. Patra$^{7}$, M. Malkan$^{14}$, H.K. Kim$^{14}$, T. Nakagawa$^{15}$, K. Malek$^{16}$,
\newauthor D. Burgarella$^{17}$, T. Ishigaki$^{18}$\\\\
$^1$ School of Physical Sciences, The Open University, Walton Hall, Milton Keynes, MK7 6AA, UK\\
$^2$ RAL Space, STFC Rutherford Appleton Laboratory, Chilton, Didcot, Oxfordshire, OX11 0QX, UK\\
$^3$ European Space Astronomy Centre, Camino Bajo del Castillo, s/n., Urb. Villafranca del Castillo, 28692 Villanueva de la Cañada, Madrid, Spain\\
$^4$ Department of Astronomy, University of Geneva, 51 Ch. Pegasi, 1290 Versoix, Switzerland\\
$^5$ South African Radio Astronomy Observatory, 2 Fir Street, Black River Park, Observatory, Cape Town 7405, South Africa\\
$^6$ Department of Physics and Electronics, Rhodes University, PO Box 94, Grahamstown 6140, South Africa\\
$^7$ Department of Pure and Applied Sciences, Midnapore City College, India\\
$^8$ ASTRON, the Netherlands Institute for Radio Astronomy, Postbus 2, 7990 AA, Dwingeloo, The Netherlands\\
$^9$ Leiden Observatory, Leiden University, PO Box 9513, NL-2300 RA Leiden, The Netherlands\\
$^{10}$ Faculty of Science Division II, Liberal Arts, Tokyo University of Science, 1-3, Kagurazaka, Shinjuku-ku, Tokyo 162-8601, Japan\\
$^{11}$ Japan Space Forum, 3-2-1, Kandasurugadai, Chiyoda-ku, Tokyo 101-0062, Japan\\
$^{12}$ Department of Earth Science Education, Kyungpook National University, Daegu 41566, Republic of Korea\\
$^{13}$ Department of Space and Astronautical Science,SOKENDAI, 3-1-1, Yoshinodai, Chuo-ku, Sagamihara, Kanagawa 252-5210, Japan\\
$^{14}$ Department of Physics and Astronomy, UCLA, Los Angeles, CA 90095-1547, USA\\
$^{15}$ Institute of Space and Astronautical Science, Japan Aerospace Exploration Agency, 3-1-1 Yoshinodai, Chuo-ku, Sagamihara, Kanagawa 252-5210, Japan\\
$^{16}$ National Centre for Nuclear Research, Pasteura 7, 02-093 Warsaw, Poland\\
$^{17}$ Laboratoire d'Astrophysique de Marseille: Marseille, Provence-Alpes-Cote d'Azu, France\\
$^{18}$ Physics Section, Iwate University, Morioka 020-8550, Japan
}

% These dates will be filled out by the publisher
\date{Accepted XXX. Received YYY; in original form ZZZ}
% Enter the current year, for the copyright statements etc.
\pubyear{2024}

\maketitle

\begin{abstract}
This paper presents a 610 MHz radio survey covering 1.94 deg$^2$ around the North Ecliptic Pole (NEP), which includes parts of the ${\it AKARI}$  (ADF-N) and Euclid, Deep Fields North. The median 5$\sigma$ sensitivity is 28~$\mu$Jy beam$^{-1}$, reaching as low as 19~$\mu$Jy beam$^{-1}$, with a synthesised beam of 3.6\arcsec $\times$ 4.1\arcsec. The catalogue contains 1675 radio components, with 339 grouped into multi-component sources and 284 'isolated' components likely part of double radio sources. Imaging, cataloguing, and source identification are presented, along with preliminary scientific results. From a non-statistical sub-set of 169 objects with multi-wavelength AKARI and other detections, luminous infrared galaxies (LIRGs) represent 66$\%$ of the sample, ultra-luminous infrared galaxies (ULIRGs) 4$\%$, and sources with L$_{IR}$~$<$~10$^{11}$ \Lsun~ 30$\%$. In total, 56$\%$ of sources show some AGN presence, though only seven are AGN-dominated. ULIRGs require three times higher AGN contribution to produce high-quality SED fits compared to lower luminosity galaxies, and AGN presence increases with AGN fraction. The PAH mass fraction is not significant, although ULIRGs have about half the PAH strength of lower IR-luminosity galaxies. Higher luminosity galaxies show gas and stellar masses an order of magnitude larger, suggesting higher star formation rates. For LIRGs, AGN presence increases with redshift, indicating that part of the total luminosity could be contributed by AGN activity rather than star formation. Simple cross-matching revealed 13 ROSAT QSOs, 45 X-ray sources, and 61 sub-mm galaxies coincident with GMRT radio sources.
\end{abstract}

\newpage

\section{Introduction}
The North Ecliptic Pole (NEP) is one of the planned targets for several current and future space telescopes. The Japanese ${\it AKARI}$  infrared satellite (\citealt{Murakami2007}) carried out a deep infrared legacy survey of this field (also known as the ${\it AKARI}$  Deep Field North (ADF-N) as discussed by \citealt{Matsuhara2006}, \citealt{takagi12}, \citealt{Matsuhara2017}). The relatively small extinction and low line of sight hydrogen column density result in relatively unobscured sight lines to the distant universe. Sensitive radio and infrared surveys have been made of the NEP  (\citep{Matsuhara2006, Matsuhara2012, Wada2008}), and the corresponding South Ecliptic Pole (SEP) field \citep{white2009, Matsuura2009, shirahata09, Matsuura2011, White2010}, with the latter paper hereafter referred to as Paper 1. These have facilitated studies of the redshift evolution of the luminosity function of radio sources,  the clustering environment of radio galaxies, the nature of obscured radio-loud active galactic nuclei (\citealt{hanami12}, \citealt{karouzos14}, \citealt{Barrufet2017a}), the association with luminous infrared galaxies (LIRGs) and ultra luminous infrared galaxies (ULIRGs), and the radio/far-infrared correlation for distant galaxies (\citealt{Barrufet2017b}).

In the present catalogue-focussed paper the results from a two-beam mosaicked GMRT radio survey at 610 MHz covering the NEP field are reported. The resulting catalogue is cross-matched with data at other wavelengths to study the source populations, as well as using the low-frequency observations of the brighter sources with the Low-Frequency Array (LOFAR) telescope at 135 MHz to recover radio spectral index information\footnote{Throughout this paper a $\rm{\Lambda}$CDM concordance cosmology was adopted with $H_0$ = 67.8 \kmsec~Mpc$^{-1}$, $\Omega$$_{\rm m}$ = 0.308, and $\Omega_{\Lambda}$ = 0.692, as estimated by the ${\it PLANCK}$  Collaboration in \cite{ade16}, and a spectral index defined as $\alpha$, where the radio flux density S$_\nu$~$\propto$~$\nu^{\alpha}$.}.

\section{Multi-wavelength observations}

In this section, a brief background overview of the panchromatic ancillary data already used in this paper is presented to provide some context (summarised in Table \ref{ancillary}). The NEP was initially observed at far-IR wavelengths in the surveys of ${\it IRAS}$  (\citealt{Hacking1987}; \citealt{aussel00}), ${\it ISO}$  (\citealt{st98}, \citealt{aussel00}),    ${\it HERSCHEL}$  (\citealt{pearson18}), COBE (\citealt{Be96}), and ${\it ROSAT}$ (\citealt{Mu2001,Mu2003}). At radio wavelengths the NEP has been observed by the VLA and JVLA (\citealt{Kollgaard1994}, \citealt{ishigaki2021} at 20 and 91 cm); Westerbork: (\citealt{Re97} at 325 MHz and \citealt{White2010} at 1.4 GHz); Effelsberg 100 m telescope at 2.7 GHz (\citealt{CB85,CB86} and \citealt{Lo88}); at optical/IR wavelengths (\citealt{Gaidos99}; and at X-ray wavelengths using ${\it ROSAT}$  (\citealt{henry01}, \citealt{Mu2001}, \citealt{henry06}). Spectroscopic observations were reported over the entire NEP-Wide field with MMT/HECTOSPEC and WYIN/HYDRA (\citealt{shim13}). Optical and infrared surveys (\citealt{hwang07}, \citealt{jeon10}, \citealt{jeon14}, \citealt{Oi2014}, \citealt{serjeant12}, \citealt{solarz15}, \citealt{Nayyeri19}, \citealt{Taylor23}) have provided key information that has been used to study the source populations of the NEP region. The ${\it AKARI}$  mission included two deep 2.4-24~$\mu$m wavelength surveys at the North Ecliptic Pole (NEP): a) covering a 0.4 deg$^2$ circular area (known as NEP-Deep - see \citealt{Matsuhara2006}); and b) a wide and shallow 2.4--24~$\mu$m survey covering a 5.8 deg$^2$ circular area surrounding the NEP-Deep field (also known as NEP-Wide - \citealt{lee09}), parts of which are covered by the GMRT radio survey. The most sensitive part of the ${\it AKARI}$  NEP Deep sky survey covered an area of 0.4 deg$^2$ with the Infrared Camera (IRC; \citealt{Onaka2007}, \citealt{Kim12}) through nine near- and mid-infrared (NIR and MIR) filters, centred at 2\mum (N2), 3\mum (N3),4\mum (N4), 7\mum (S7), 9\mum (S9W), 11\mum (S11),15\mum (L15), 18\mum (L18W), and 24\mum (L24) as described by \citealt{Onaka2007, takagi10, Kim12, pearson10, pearson18}, and recently summarised by \citealt{shim22}.
\begin{table*}
\vspace{0pt}
%\begin{table*}
\caption{Summary of ancillary observations available for the GMRT NEP deep field referred to from this paper. AB magnitude is used throughout this Table unless otherwise indicated.}
\begin{scriptsize}
\fontsize{8}{10}\selectfont
\begin{tabular}{l c c c l l}
\hline
\multicolumn{1}{l}{Wavelength} & \multicolumn{1}{c}{Telescope/instrument} & \multicolumn{1}{c}{Area sq. degrees} & \multicolumn{1}{c}{Beam size} & \multicolumn{1}{c}{Depth 5$\sigma$} & \multicolumn{1}{c}{Data set}\\ 
\multicolumn{1}{l}{(1)} & \multicolumn{1}{c}{(2)} & \multicolumn{1}{c}{(3)} & \multicolumn{1}{c}{(4)} &  \multicolumn{1}{c}{(5)} &  \multicolumn{1}{c}{(5)}  \\
\hline
FUV, NUV & ${\it GALEX}$  & 1 & 6\arcsec & 26.1-26.7 mag &\cite{karouzos14}\\
$u, g, r, i, z$ & CFHT & 2 & 1\arcsec & 24 mag & \cite{hwang07}\\
$B, R, I$ & Maidanak/SNUCAM & 4 & 1\arcsec & 22.3-23.5 mag& \cite{jeon10}\\ 
Optical spectroscopy & WIYN/HYDRA & 5 & 1\arcsec & $R$ $\leq$~21.5 mag & \cite{shim13}\\ 
Optical spectroscopy & MMT/HECTOSPEC & 5 & 1\arcsec & $R$ $\leq$~21.5 mag & \cite{shim13}\\ 
$J, H$& KPNO 2.1m/FLAMINGOS & 5.2 & 1\arcsec & 21.4 mag & \cite{jeon14}\\ 
2, 3, 4~$\mu$m & ${\it AKARI}$ /IRC & 5.4 & 4.2-5.5\arcsec & 20.9-21.1 mag & \cite{Kim12}\\ 
3.4-22~$\mu$m & ${\it WISE}$  & All Sky & 6-12\arcsec & 19.9-14.6 mag & \cite{wright10}\\ 
7, 9, 11~$\mu$m & ${\it AKARI}$ /IRC &5.4&2.9-3.3\arcsec &19-19.5 mag & \cite{Kim12}\\ 
15, 18 ~$\mu$m & ${\it AKARI}$ /IRC&5.4& 4.6\arcsec&18.6-18.7 mag & \cite{Kim12}\\
24 ~$\mu$m & ${\it AKARI}$ /IRC&5.4 & 6.7\arcsec&18.0 mag & \cite{Kim12}\\
65, 90, 140, 160~$\mu$m & ${\it AKARI}$ /FIS & All Sky  & 37-50\arcsec & 50~mJy (90~$\mu$m)&\cite{yamamura10}\\ 
250, 350, 500~$\mu$m & ${\it HERSCHEL}$ /SPIRE & 9 & 18-36\arcsec & $\sim$  37.5-54 mJy &\cite{pearson17}\\
100, 160~$\mu$m &   ${\it HERSCHEL}$ /PACS & 0.67 & 5-12\arcsec & $\sim$ 20-50 mJy&\cite{pearson17}\\
850~$\mu$m & JCMT/S2CLS & 0.6 & 15\arcsec & $\sim$ 6 mJy & \cite{geach17}\\
20 cm & WSRT & 1 & 15\arcsec & 105~$\mu$Jy& \cite{White2012}\\
\hline
 \end{tabular}
\end{scriptsize}
\label{ancillary}
\end{table*}

\section{Observations and data analysis}
\label{section:observations}
\subsection{GMRT observations}
The NEP was observed in 2007 and 2009 as part of a two-beam mosaic with the Giant Meterwave Radio Telescope (GMRT). The GMRT has thirty fully steerable 45-metre diameter telescopes spread over a 25 km area in southeast India. Sixteen of the telescopes are arranged in a Y-shaped array, with the remaining fourteen randomly distributed within a one-kilometre area.  Full details of the GMRT characteristics and performance are described in \citet{sw1}. 

The initial observations were made at the pointing centre RA (J2000) 17$^{\rm h}$55$^{\rm m}$24$^{\rm s}$, Dec (J2000) +66$^{\circ}$37$'$32$''$ for 12 hours on the 2nd and 3rd August 2007 (GMRT Proposal ID's 12SSC01 and 15SSC01, PI S. Serjeant, Archive Observations 3288, 3291, 4160, 4162). Further observations were made at the offset pointing centre 18$^{\rm h}$02$^{\rm m}$03$^{\rm s}$, 66$^{\circ}$31$'$55$''$ on 26th and 27th January 2009 for 17 hours, and the final data set was formed from a mosaicked image made from these two independent pointings. The central frequency of the observations was 610 MHz, with a bandwidth of 32 MHz. The observations were taken in the so-called Indian Polar mode, with the visibilities from both polarisations of all 30 antennas measured with 128 x 125 kHz spectral channels, which efficiently provides the full sensitivity of the GMRT, whilst minimising bandwidth smearing effects. The data were collected using a short visibility integration time with the correlator (16.9 seconds in August, and 4.2 seconds in January) to facilitate the identification of radio frequency interference (RFI) as well as minimising smearing effects.

\begin{figure*}
 \centering
% \begin{center}
\includegraphics[width=0.7\linewidth,angle=270]{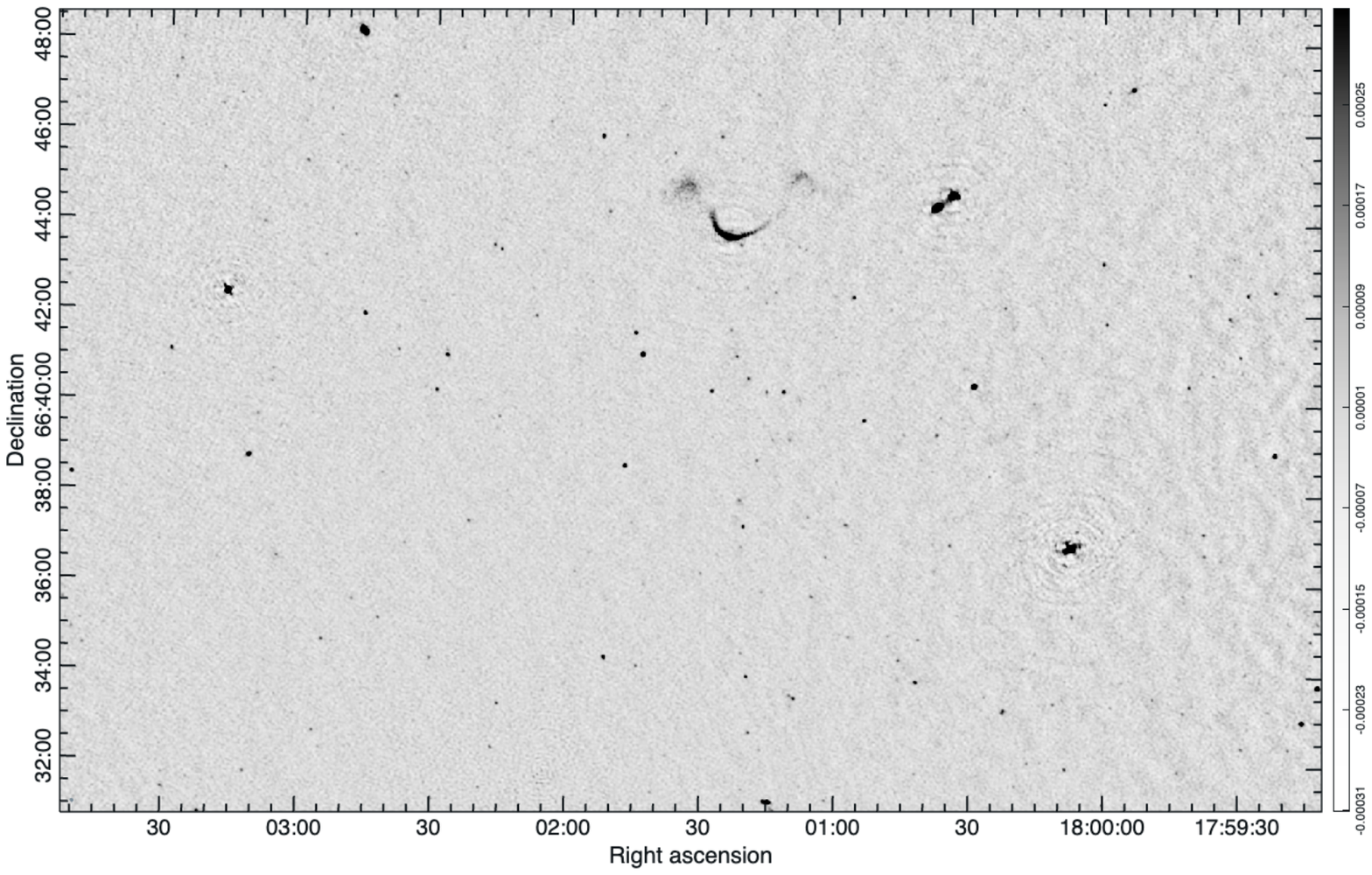}\\
% \end{center}
 \caption{A 17\arcmin x 28\arcmin section of the GMRT 610 MHz two-beam mosaic map, corrected for primary beam response, and shown with a squared greyscale stretch, and the shaded bar on the right showing units of Jy per beam.}
 \label{centralarea}
\end{figure*}

The data were reduced using AIPS++\footnote{http://aips2.nrao.edu}, with corrections made for the system temperature ($T_{sys}$), which varied with antenna, ambient temperature, and elevation. Additionally, data for antennae with high errors in antenna-based solutions were flagged over certain time ranges, along with a few baselines flagged based on closure errors on the bandpass calibrator. Channel and time-based flagging of data points corrupted by RFI was applied using a median filter with a 6$\sigma$ threshold, with residual errors above 5$\sigma$ also being flagged after several rounds of imaging and self-calibration.

The radio sources 3C 48 (29.43 Jy) and 3C 286 (21.875 Jy) were used as the primary bandpass and flux density calibrators, with secondary reference sources J1634+627 (9.16 Jy) and J1400+621 (5.96 Jy) used as phase calibrators. After first applying bandpass, gain and phase corrections to the phase calibrators, the flux density, bandpass, gain and phase calibrations were then applied to the target field. The final images of each of the field centres were made after several rounds of phase self-calibration, followed by a single round of amplitude self-calibration, and normalisation by the median gain for all of the data. These were then combined, correcting for the primary beam response, and mosaicking in the image plane weighted by the individual variance maps. This final image had a total bandwidth of 32 MHz, and a synthesised half-power beam of 3.6\arcsec $\times$ 4.1\arcsec. A small area from the most sensitive part of the mosaicked area is shown in Figure \ref{centralarea}, and the noise map from the whole area covered is shown in Figure \ref{surveyarea}. 

The theoretical sensitivity is given by:
\begin{equation}
\sigma = \frac{\sqrt{2}~T_{sys}}{G\sqrt{n(n-1)~f_d~\Delta \nu~\tau }}
\label{Eqn:noiseequn}
\end{equation}
where $T_{sys} \sim$ 92$\thinspace$K, the antenna gain $G \sim$0.32K Jy$^{-1}$, $n$ is the number of working antennas, $\Delta \nu$ is the bandwidth and $\tau$ is the total duration of the observations, $f_d$ is the total fraction of data which has been used for making the final image. The expected noise at peak response in the final two-beam mosaic was $\sim$18.9~$\mu$Jy beam$^{-1}$, which was within a few per cent of that measured in the most sensitive region of the final mosaicked map of $\sim$19.0~$\mu$Jy beam$^{-1}$.
\subsection{LOFAR observations}

To estimate the radio spectral index of the GMRT 610 MHz detections, data were analysed from an archival 8-hour observation of the NEP field in 2013 (LOFAR Cycle 1 observing period) with the LOFAR High Band Antenna (HBA) at frequencies from 114-156 MHz effective centre frequency of 135 MHz HBA observation L158162).

A much deeper and wider area LOFAR HBA survey of the Euclid Deep Field North, is currently still being observed with LOFAR (\citealt{bondi23}) as part of its Deep Surveys campaign, and so we defer publishing a full LOFAR HBA catalogue in this paper based on the 2013 LOFAR data until observations of the pending deep data set have been completed. However, we use this early LOFAR HBA data to estimate the spectral indices of the brighter source population not requiring the full LOFAR sensitivity. The LOFAR HBA data were processed through the standard LOFAR pipeline as fully described in \citet{Shimwell2019} and \citet{Wi2019}. After imaging, a component catalogue was extracted from the primary beam corrected image using the PyBDSF~\footnote{ https://github.com/lofar-astron/PyBDSF} source finder, to provide consistency with the GMRT component extraction. The synthesised beam size for the LOFAR data was 6\arcsec, positional accuracy 0.2\arcsec $\thinspace$and intensity calibration better than 20$\%$. There is excellent agreement between this small sub-set used here and the larger LOFAR data currently being prepared for publication.

\section{Calibration}
\label{Calib}

In the following sections, details of the GMRT calibration and data reduction methodology are presented. Since much of this is in common with our recent North and South Ecliptic Pole radio surveys using the WSRT and ATCA telescopes, the reader is referred to these earlier papers (see Paper 1 and \citealt{White2012}). Here we focus only on those calibration issues specific to the GMRT and LOFAR data in this paper.

\subsection{Flux accuracy and error estimates}
The GMRT data were corrected for various instrumental effects including a) the primary beam response of the antenna elements (see Section \ref{Primary}); b) chromatic aberration or bandwidth smearing effects resulting from the finite bandwidth \citep{br89,co89} (see also Section \ref{Smearing}) c) time-average smearing due to the finite integration time (Section \ref{Time}); and d) incompleteness at low signal to noise levels (Section \ref{Completeness}). The approach adopted to tackle these issues is described in the following sub-sections.

\subsection{Primary beam response of antenna elements}
\label{Primary}
\begin{figure}
 \centering
% \begin{center}
 \includegraphics[width=0.7\linewidth,angle=270]{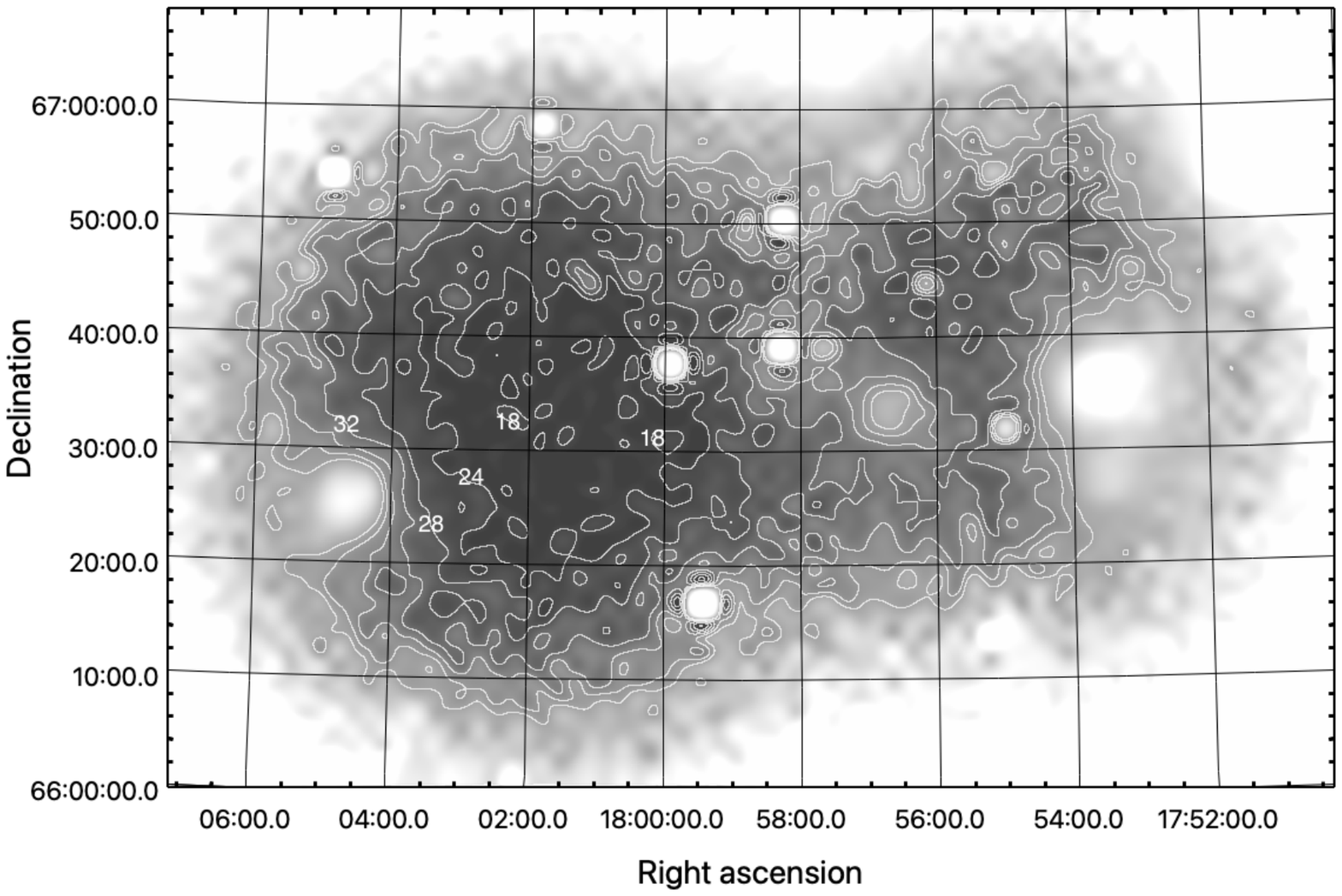}\\
% \end{center}
 \caption{The contours show the rms noise levels in the final mosaicked image in units of~$\mu$Jy beam$^{-1}$, estimated locally from the noise map by binning the data into 40~$\times$~40 pixel regions. The lowest contour displayed (close to RA 18$^h$ 02$^m$ and Dec +66$^{\circ}$ 26$^m$) is 18~$\mu$Jy per beam, with the subsequent contours increasing at intervals of 4~$\mu$Jy per beam, with several of the contours marked. The greyscale image is illustrative to guide the eye to the most sensitive parts of the map (marked in black), whilst saturated white regions in the map indicate areas where residual calibration inaccuracies and/or bright sources have increased the effective noise level, or at the edges of the mapped area where the primary beam attenuation becomes significant. This variation in the sensitivity and survey depth has been taken into account in later sections, including in the component extraction and source catalogue.}
 \label{surveyarea}
\end{figure}

The sensitivity of the primary beam changes as a function of distance from the pointing centre and is represented as the primary beam response $B(x)$. An 8th-order polynomial was fitted to the primary beam response and applied to the data. The final images of both fields were individually corrected for the FWHM = 43$^{\prime}$$\thinspace$ primary beam response, which is approximated by:
\begin{equation}
B(x) = 1 + \frac{a}{10^3} x^2 + \frac{b}{10^7} x^4 + \frac{c}{10^{10}} x^6 + \frac{d}{10^{13}} x^8
\end{equation}
where $x$ is proportional to the square of the distance from the pointing position in units of [arc min $\times$ freq (GHz)]$^2$ and $a$, $b$, $c$ and $d$ are coefficients of the polynomial fit with values, $a = -3.486$, $b= 47.749$, $c= -35.203$ and $d = 10.399$ respectively \citep{KR01}. The two individual primary beam-corrected fields were then mosaicked in the image plane after accounting for the variances across their fields. The synthesised beam in the final mosaicked image was 3.6\arcsec $\times$ 4.1\arcsec~FWHM estimated from measurements of isolated point sources. The final image has a median root mean square (rms) noise level averaged across the whole field of 28~$\mu$Jy beam$^{-1}$ after primary beam correction, although in a few areas near bright sources, the noise is greater due to calibration issues. Figure \ref{centralarea} shows a cut out of the central part of the mosaic, and Figure \ref{surveyarea} shows a map of the rms noise level corrected for primary beam attenuation. The highest sensitivity close to RA (J2000) 18$^{\rm h}$ 02$^{\rm m}$ and Dec (J2000) +66$^{\rm \circ}$ 26$^{\rm m}$ is 19.0~$\mu$Jy per beam, and a similar secondary minimum close to RA 17$^{\rm h}$ 56$^{\rm m}$ and Dec +66$^{\rm \circ}$ 41$^{\rm m}$.

\subsection{Astrometric accuracy}
\label{positionalaccuracy}

The rms positional uncertainty in each of the measured sky coordinates ($\Delta$RA or $\Delta$Dec) of a point source is primarily determined by noise fluctuations, and by errors in the phase measurement. For an rms brightness fluctuation $\sigma$ and FWHM resolution $\theta$, the positional fluctuation ${\it \sigma_p}$ is given by (from \citealt{Re97}):
\begin{equation}
{\sigma}_p \approx \frac{\sigma \theta}{2 S_{peak}}
\label{positionaccuracy}
\end{equation}
Systematic errors can also affect the overall astrometric accuracy, and it has become common practice, particularly at low radio frequencies, to compare the measured positions with those determined from compact extragalactic sources visible from other surveys. In making such comparisons it is important to consider how the contribution of the astrometric errors from these other radio surveys on cross-matching and inter-comparison. 

\begin{figure}
 \centering
% \begin{center}
\hspace{-1.5cm}
 \includegraphics[width=1.2\linewidth,angle=0]{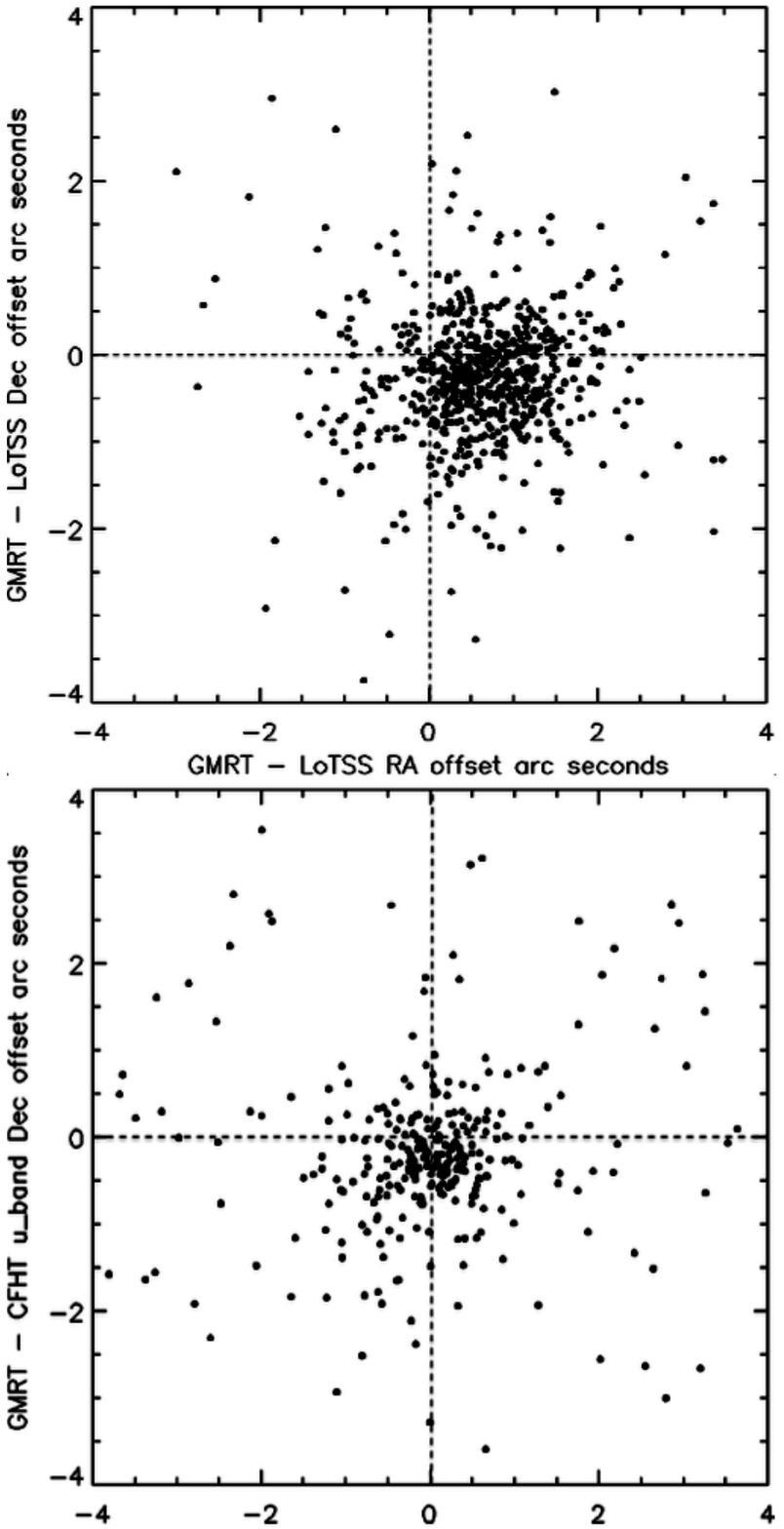}
 % \hspace{-1cm}
 %\includegraphics[width=0.79\linewidth,angle=0]{GMRT_CFHT_position_offsets.pdf}\\
% \end{center}
 \caption{[Left] The measured positional offsets in arc seconds between bright compact GMRT and LOFAR High Band Array components for which the coordinate frame is established using methods described by \citealt{Shimwell2019}. [Right] The measured positional offsets in arc seconds between the bright compact GMRT components and bright compact (CFHT $u$-band brighter than 24.5 magnitudes) sources, where the CFTH reference frame is as described by \citet{huang20}. The LOTSS detections appear to be systematically offset from the GMRT positions by (0.5, -0.5)\arcsec.}
 \label{gmrt_lofar}
\end{figure}

Although the NRAO VLA Sky Survey (NVSS), the 1.5 GHz VLA-NEP survey of \citet{Kollgaard1994} and the GMRT TGSS survey (\citealt{intema2017}) share some common spatial overlap with the GMRT NEP field, the positional errors of these surveys were insufficient for use in establishing the GMRT reference frame due to their different angular resolution. We instead chose to compare the GMRT with LOFAR HBA data discussed earlier in this paper, which has an accuracy of 0.2\arcsec~(\citealt{Shimwell2019}). The resulting GMRT-LOFAR positional offsets are shown in Figure \ref{gmrt_lofar}, where we note a small, systematic shift between the two reference frames.

The GMRT radio component positions were also compared with a recent high-resolution radio catalogue using the JVLA telescope at the L band (1.0--2.0 GHz) with a 1.52\arcsec ~$\times$ 1.21\arcsec~synthesised beam (\citealt{ishigaki2021}). From cross-matching with the GMRT and JLVA data, 83 radio components were co-located within 0.5\arcsec~of each other, and 95$\%$ of the sources within 1\arcsec~of each other, with the two independent astrometric solutions agreeing to within about 0.5\arcsec.

As a final check, a sample of bright, compact, optical galaxy nuclei and known AGN, which have well-determined optical positions of 1\arcsec~or less, were cross-matched with compact isolated radio components (defined as discussed in Section \ref{detection}), the ${\it ALLWISE}$  source catalogue\footnote{http://wise2.ipac.caltech.edu/docs/release/allwise/}, the \citet{shim13} optical spectroscopic survey, and the $u$-band CFHT optical survey of \citet{huang20}. The uncertainty in adopting this type of approach depends on the radio components being coincident with the optical or infrared objects, but they are consistent with the radio-radio cross-match analysis. This analysis showed an offset of about 0.2\arcsec~between bright optical point sources and compact GMRT components, with 95$\%$ of the cross-matched sources being within a distance of 1.3\arcsec~of the GMRT compact component position - as also shown in Figure \ref{gmrt_lofar} where the positional offsets to bright ($u$ $<$ 24 magnitude) compact optical detections from the CFHT survey are plotted.

Based on these checks of the astrometric efficacy, a conservative matching radius of 4\arcsec~has been adopted in this paper, although it is noted that most of the direct positional cross-matches with the most accurate comparator, the JVLA data, are all found to be co-located within a radius of 1\arcsec~of the compact GMRT radio components. The positional errors reported in Table \ref{sourcecatalogueshort} are those directly estimated from the GMRT data alone. The number of false matches that resulted from artificially adding offsets of $\pm$ 5\arcmin~to the north, south, east and west of the nominal pointings was also examined, which suggested that the identification efficiency was $\approx$~98$\%$. Although sub-arcsecond positional accuracy is normally needed to make firm associations, the adopted cross-matching strategy was felt to be the best that could be done with the current calibration strategy, and sub-arcsecond cross-matching to the JVLA and future LOFAR observations is left for a more sophisticated optical/IR/radio reference frame.

\subsection{Bandwidth Smearing}
\label{Smearing}
Bandwidth smearing is the radio analogue of optical chromatic aberration, affecting all synthesis observations made with finite channel width. The resultant smearing of an image is proportional to the fractional bandwidth $\Delta \nu / \nu_0$ (where $\Delta \nu$ is the bandwidth and $\nu _0$ is the central frequency of observation) and to the distance from the phase centre in units of the synthesised beam (e.g. \citealt{Th01}). For the final mosaic image, the maximum flux reduction due to bandwidth smearing was estimated to be less than 1 per cent, without no significant variation with the position in the mosaic.

\subsection{Time-average smearing}
\label{Time}
For long integration times, time-average smearing elongates point sources away from the centre of the map in the azimuthal direction \citep{bschwab99}. The integration time of the data taken from the NEP on the 2nd and 3rd of August 2007 was 16.908 s and on the 26th and 27th of January 2007 was 4.227 s. Based on experience from the GMRT, the effect on the reduction of point source fluxes is expected to be no more than $\sim$1$\%$ for a point source located 22\arcmin ~from the field centre of each of the pointing positions, and should not significantly affect the measured source sizes.

\subsection{Clean bias}
Flux density measurements in radio surveys can be affected by a clean bias effect, where a systematic underestimation of the total and peak fluxes of the source is a consequence of the redistribution of the flux from point sources to noise peaks in the image \citep{Be95,Wh97}. Clean bias also depends on the dynamic range of the primary beam (peak to brightest side lobes) and the uv-plane coverage of the observations. The GMRT observations had very good uv-coverage, so following criteria suggested by \cite{Ga00}, the cleaning limit was set at 5 times the theoretical noise to mitigate against this effect, and it is therefore considered insignificant.

\subsection{Resolution bias}
Resolution bias is an effect in which the peak flux densities of weak extended sources fall below a given detection limit, yet still, have total integrated flux densities above the survey limit. The same methodological approach described in Section 4.4 of Paper 1 has been followed, and it is estimated that any flux reduction due to smearing effects for the GMRT data is $\le$ 3$\%$.

\subsection{Summary of flux density corrections for systematic effects}
Summarising the previous subsections, various systematic effects described above in Section \ref{Calib} have been used to make an empirical estimation the GMRT flux densities. The flux densities reported in Table \ref{sourcecatalogueshort} have been corrected for the various effects described as follows (see also \citealt{prandoni10a}):
\begin{equation}
S_{corr}  = \frac{{S_{meas}}}{{k\,\left[ {a\;{\rm log}\left( {\frac{{S_{meas}}}{\sigma }} \right) + b} \right]}}
\label{prandonicorr}
\end{equation}
where $S_{meas}$ is the flux measured in the GMRT images (reported in the source catalogue in this paper). The parameter ${\it k}$ represents the smearing correction. This has a value of unity (i.e. no correction) when the equation is applied to integrated flux densities and $\le$ 1 when dealing with peak flux densities. By comparing the peak and integrated intensities of the GMRT sources (see Section 4.4 of \citet{White2010} for a more general discussion on this) it is estimated that ${\it k}$ = 0.92 (i.e. an 8$\%$ smearing effect which redistributes flux and reduces the peak fluxes). 

\subsection{Completeness and reliability}
\label{Completeness}
The false detection rate for the GMRT survey was estimated from an identical source extraction on a flux-inverted copy of the image, which was then beam-corrected. This inverted map had similar background properties and should provide an equal number of false detections as on the real image. Fifteen compact sources were detected from the inverted image after correcting for the major artefacts close to bright sources, with fluxes down to 90~$\mu$Jy. Visual inspection of each of these detections failed to show any clear association with the bright source artefacts or location relative to the primary beam centres. Therefore we estimate that the total false detection rate for the GMRT data using our extraction criteria is $\le$~0.9$\%$.

\section{Source Catalogue}
\label{detection}
\begin{figure}
 \centering
 \hspace*{-0.5cm}
 \includegraphics[angle=-90, width=1.1\linewidth]{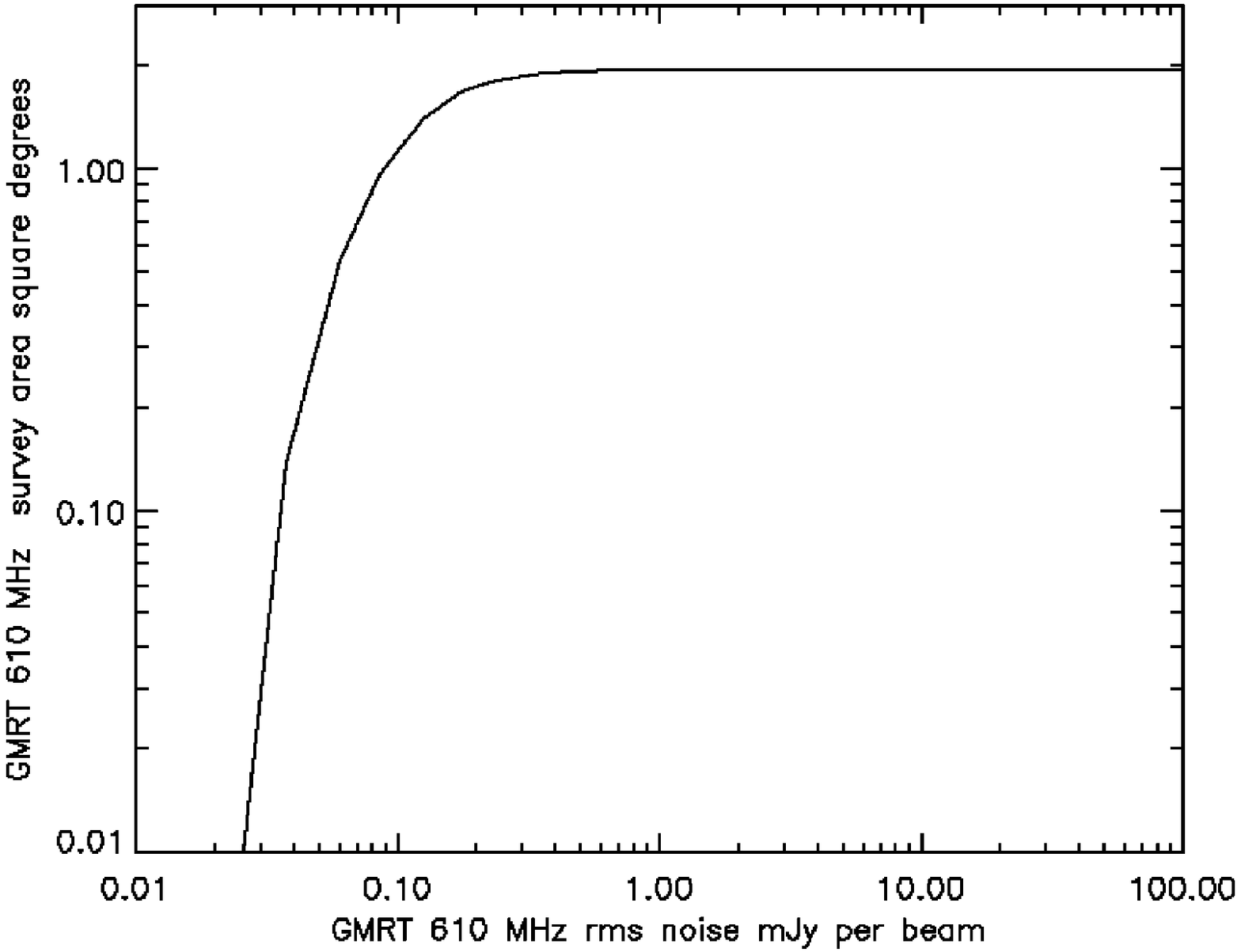}\\
 \caption{The horizontal axis shows the rms detection threshold as a function of areal coverage over the maximum of 1.94 square degrees used for the catalogue, with one square degree having an rms sensitivity lower than 0.1 mJy per beam, and 0.45 square degrees lower than 50~$\mu$Jy per beam}
 \label{area}
\end{figure}

In this section, we describe the methodology used in creating the source component catalogue. Although the mosaicked region offers both wide-field coverage and good sensitivity, it has a non-uniform noise distribution which needs to be taken into account. Source detection in this circumstance is best performed using locally determined noise levels measured from the noise map shown in Figure \ref{area} \citep{hopkins98, White2010}.

For source extraction and cataloguing the PyBDSF was adopted. PyBDSF initially builds a noise map from the pixel data using variable mesh boxes to deal with regions having higher noise around the brightest sources. A 5$\sigma$ peak noise threshold was then used to define sources using an island threshold of 4$\sigma$ as a boundary using the wavelet-decomposition option.

An approach to grouping radio components together is included in the PyBDSF source extractor that attempts to associate the recovered Gaussian components into islands of emission. Initially, the group of Gaussians is calculated as having several distinct peaks (defined from a pre-set pixel threshold) of emission (i.e. that have negative gradients in 8 directions). These peaks are CLEANed assuming the theoretical beam using the beam parameters stored in the image fits file header. If the rms uncertainty of the residual sub-image exceeds that of a nearby source-free region (determined locally from the image), the maximum pixel in the residue is greater than a 5$\sigma$ threshold for the locally estimated rms, and separated by at least 0.5 beams (and$\sqrt{5}$ pixels) from all previous peaks, then this residual peak is identified as a new Gaussian peak, and is added to an island. Groups of Gaussians were considered to be a part of the same island if a) no pixel on the line joining the centres of any pair of Gaussians had a value less than the preset island threshold (which was set at 4$\sigma$ against the local rms value), and b) the centres were separated by a distance less than half the sum of their FWHMs along the line joining them. Once the Gaussians that belong to a multi-component source were identified, fluxes for the grouped Gaussians were summed to obtain the total flux. The uncertainty of the total flux was estimated by taking the square root of the sum of the uncertainties on the total fluxes of the individual Gaussians. In this case, the estimated source Right Ascension and Declination were set to the source centroid determined from moment analysis, which refers to the statistical method used to determine the centroid (position) and its uncertainties for detected sources. This method involves calculating the moments of the pixel intensity distribution within a detected source. In the source catalogue below the parameters of the individual Gaussian components are listed, and discussion of the issue of multi-component sources is deferred until Section \ref{complexsources}.

The final PyBDSF catalogue contained 1675 radio components, of which 339 were automatically grouped into multi-component sources. A further selection of double-lobed sources from the remaining single Gaussian sources using criteria described in Section \ref{complexsources}, resulted in a further 284 of the remaining single Gaussian components additionally being classified as double sources. As a final check of this semi-automated classification, all double and multi-component sources were also examined for possible optical, radio and/or infrared associations close to the centroid of the position between two adjacent radio components, as an additional way to identify possible FRII radio associations. All of the catalogued radio components were visually inspected to confirm that the classification methodology appeared to be correct, and did not miss any that might otherwise not be identified (such as multiple-lobed sources, where the associated galaxy is often close to the centroid between the radio components).

A sample from the final source catalogue is displayed in Table~ \ref{sourcecatalogueshort}, with the complete catalogue listed in the electronic online version of this paper.
\begin{table*}
\vspace{0pt}
%\begin{table*}
\caption{The complete source catalogue (the full version is available as Supplementary Material in the online version of this article). The source parameters listed in the catalogue are: (1) a short form running number, (2) the source Right Ascension referenced from the self-calibrated reference frame and (3) the uncertainty in arc seconds, (4) the Declination (J2000)  and (5) the uncertainty in arc seconds, (6) the peak flux density, S$_{\rm peak}$, (7) its associated rms uncertainty, (8) the integrated flux density, S$_{\rm total}$ and (9) the associated uncertainty, (10) the major axis full width at half maximum in arc seconds of the fitted Gaussian source profile, (11) the minor axis full width at half maximum in arc seconds of the fitted Gaussian source profile and (12) position angle in degrees measured east of north. For these last three the orientation (major and minor axes full width at half maximum in arc seconds, and the position angle in degrees measured east of north. Sources constituting the smaller 169 source sample referred to later in the paper in Section \ref{169Sample} are indicated with a star following their short form running number in column 1.}
\begin{scriptsize}
\fontsize{8}{10}\selectfont
\begin{tabular}{l l r r r r r r  r r r r}
\hline
\multicolumn{1}{l}{No} & \multicolumn{1}{c}{RA} & \multicolumn{1}{c}{$\Delta$RA} & \multicolumn{1}{c}{DEC} &  \multicolumn{1}{c}{$\Delta$DEC} & \multicolumn{1}{c}{S$_{\rm peak}$} & \multicolumn{1}{c}{$\Delta$S$_{\rm peak}$} & \multicolumn{1}{c}{S$_{\rm total}$} & \multicolumn{1}{c}{$\Delta$S$_{\rm total}$} & \multicolumn{1}{c}{$\theta_{maj}$} & \multicolumn{1}{c}{$\theta_{min}$} & \multicolumn{1}{c}{$PA$}\\ 
 \multicolumn{1}{l}{} & \multicolumn{1}{c}{h:m:s.s} & \multicolumn{1}{c}{${\prime\prime}$} & \multicolumn{1}{c}{d:m:s.s} & \multicolumn{1}{c}{${\prime\prime}$} & \multicolumn{1}{c}{mJy} & \multicolumn{1}{c}{mJy} & \multicolumn{1}{c}{mJy}  & \multicolumn{1}{c}{mJy}  & \multicolumn{1}{c}{$^{\prime\prime}$} &  \multicolumn{1}{c}{$^{\prime\prime}$} &  \multicolumn{1}{c}{$\ensuremath{^\circ}\,$}\\
  \multicolumn{1}{l}{} & \multicolumn{1}{c}{} & \multicolumn{1}{c}{} & \multicolumn{1}{c}{} & \multicolumn{1}{c}{} & \multicolumn{1}{c}{beam$^{-1}$} & \multicolumn{1}{c}{beam$^{-1}$} & \multicolumn{1}{c}{}  & \multicolumn{1}{c}{}  & \multicolumn{1}{c}{} &  \multicolumn{1}{c}{} &  \multicolumn{1}{c}{}\\
\multicolumn{1}{l}{(1)} & \multicolumn{1}{c}{(2)} & \multicolumn{1}{c}{(3)} &  \multicolumn{1}{c}{(4)} & \multicolumn{1}{c}{(5)} & \multicolumn{1}{c}{(6)} & \multicolumn{1}{c}{(7)} & \multicolumn{1}{c}{(8)} & \multicolumn{1}{c}{(9)} & \multicolumn{1}{c}{(10)}  & \multicolumn{1}{c}{(11)} & \multicolumn{1}{c}{(12)} \\
\hline
  57 & 17:51:13.54 & 0.61 & +66:28:03.3 & 0.8 & 888.0 & 118.2 & 342.0 & 88.7 & 6.8 & 5.6 & 161.3\\
  58 & 17:51:19.33 & 0.43 & +66:02:04.7 & 0.6 & 3835.1 & 658.9 & 2232.3 & 438.9 & 6.5 & 3.9 & 31.7\\
  59* & 17:51:20.82 & 0.23 & +66:33:59.9 & 0.2 & 502.2 & 111.1 & 492.0 & 63.5 & 4.5 & 3.3 & 39.9\\
  60 & 17:51:22.12 & 0.04 & +66:36:38.2 & 0.0 & 2767.6 & 112.2 & 2548.4 & 66.6 & 4.0 & 4.0 & 132.4\\
  61 & 17:51:23.93 & 0.05 & +66:57:14.8 & 0.0 & 6722.3 & 211.6 & 5203.6 & 132.1 & 4.8 & 4.0 & 111.6\\
  62 & 17:51:31.78 & 0.26 & +66:38:21.2 & 0.4 & 326.4 & 112.4 & 347.6 & 62.3 & 4.4 & 3.2 & 25.6\\
  63 & 17:51:31.89 & 0.17 & +67:05:32.5 & 0.2 & 1724.5 & 320.2 & 1793.8 & 182.3 & 3.8 & 3.7 & 86.1\\
  64 & 17:51:33.42 & 0.27 & +66:01:52.1 & 0.3 & 1849.2 & 693.9 & 2207.7 & 373.1 & 3.9 & 3.1 & 45.1\\
  65 & 17:51:33.97 & 0.34 & +66:31:56.2 & 0.3 & 360.2 & 113.1 & 359.0 & 64.9 & 4.3 & 3.5 & 60.7\\
  66 & 17:51:37.38 & 0.29 & +66:31:23.6 & 0.2 & 937.2 & 115.4 & 611.8 & 75.7 & 5.3 & 4.3 & 105.2\\
  67 & 17:51:37.38 & 0.14 & +66:31:22.3 & 0.1 & 533.4 & 124.8 & 729.5 & 63.9 & 3.6 & 3.0 & 59.3\\
\hline
 \end{tabular}
\end{scriptsize}
\label{sourcecatalogueshort}
\end{table*}
\subsection{Component extraction}
 \label{complexsources}
\begin{figure*}
 \centering
% \begin{center}
%\vspace*{-1.5cm}
%\includegraphics[width=0.33\linewidth,angle=0]{complex1.pdf}
%\includegraphics[width=0.33\linewidth,angle=0]{complex2.pdf}
%\includegraphics[width=0.33\linewidth,angle=0]{complex3.pdf}\\
%\vspace*{-3.cm}
%\includegraphics[width=0.33\linewidth,angle=0]{complex4.pdf}
%\includegraphics[width=0.33\linewidth,angle=0]{complex5.pdf}
%\includegraphics[width=0.33\linewidth,angle=0]{complex6.pdf}\\
%\vspace*{-3.cm}
%\includegraphics[width=0.33\linewidth,angle=0]{complex7.pdf}
%\includegraphics[width=0.33\linewidth,angle=0]{complex8.pdf}
%\includegraphics[width=0.33\linewidth,angle=0]{complex9.pdf}\\
%\vspace*{-3.cm}
%\includegraphics[width=0.33\linewidth,angle=0]{complex10.pdf}
%\includegraphics[width=0.33\linewidth,angle=0]{complex13.pdf}
%\includegraphics[width=0.33\linewidth,angle=0]{complex12.pdf}\\
%\vspace*{-1.cm}
\includegraphics[width=0.9\linewidth,angle=0]{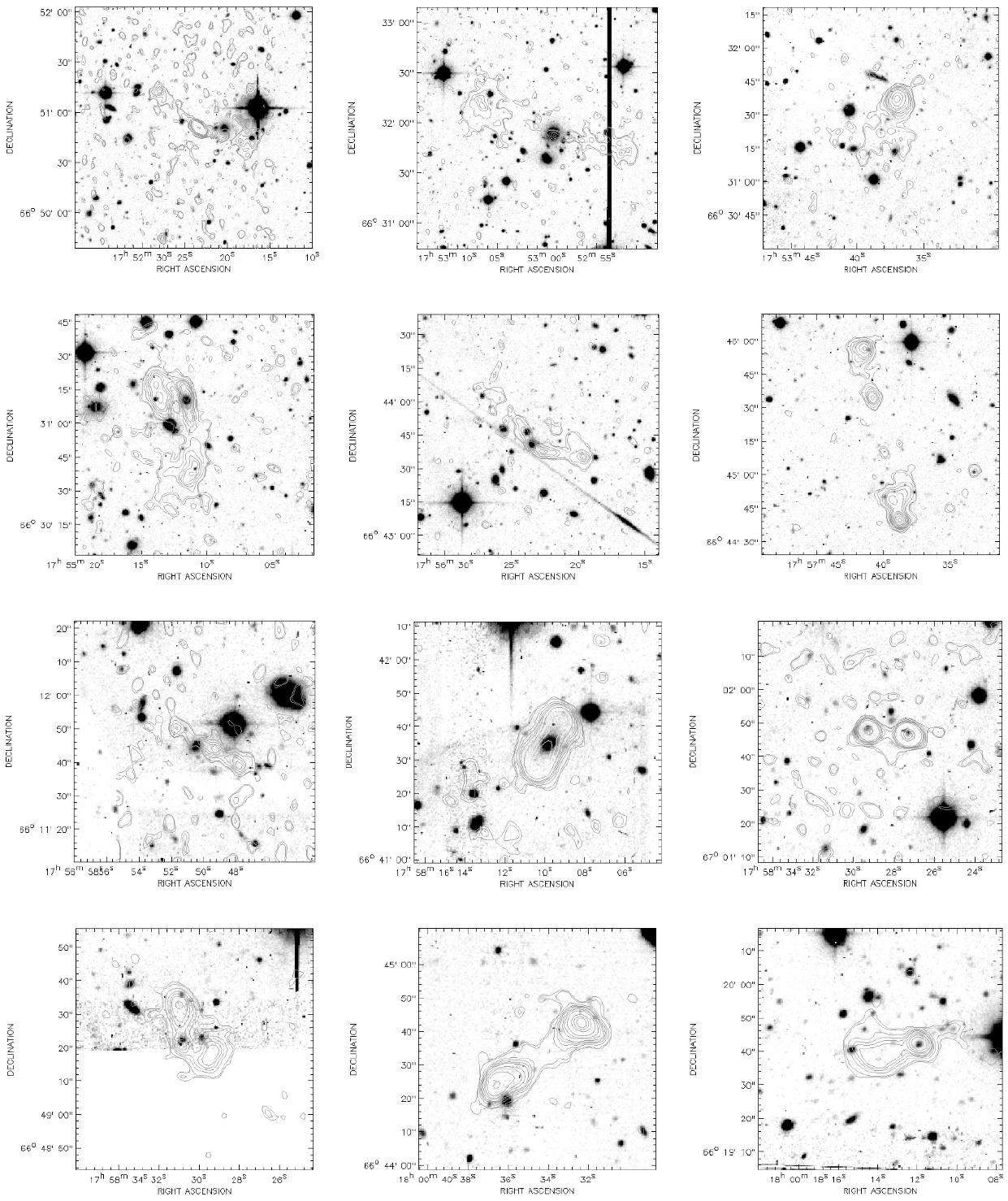}\\
 \caption{Cut-out maps around multi-component sources. The displayed contour levels showing the radio emission and grey scalings showing the optical emission are arbitrary and linearly spaced, and chosen to show the bright compact components in extended sources, as well as any faint extended surrounding emission, the local noise levels, and in some cases nearby fainter sources. For accurate fluxes of the individual components, the reader is referred to Table \ref{sourcecatalogueshort}. The black stripes and missing optical data are seen in some of the images due to the physical arrangement, and charge transfer effects in the CCD chips in the background SUBARU R-band images.}
 \label{extendedsources1}
\end{figure*}

\begin{figure*}
 \centering
% \begin{center}
%\vspace*{-1.5cm}
%\includegraphics[width=0.33\linewidth,angle=0]{complex16.pdf}
%\includegraphics[width=0.33\linewidth,angle=0]{complex14.pdf}
%\includegraphics[width=0.33\linewidth,angle=0]{complex15.pdf}\\
%\vspace*{-3.cm}
%\includegraphics[width=0.33\linewidth,angle=0]{complex17.pdf}
%\includegraphics[width=0.33\linewidth,angle=0]{complex18.pdf}
%\includegraphics[width=0.33\linewidth,angle=0]{complex19.pdf}\\
%\vspace*{-1.cm}
\includegraphics[width=0.7\linewidth,angle=270]{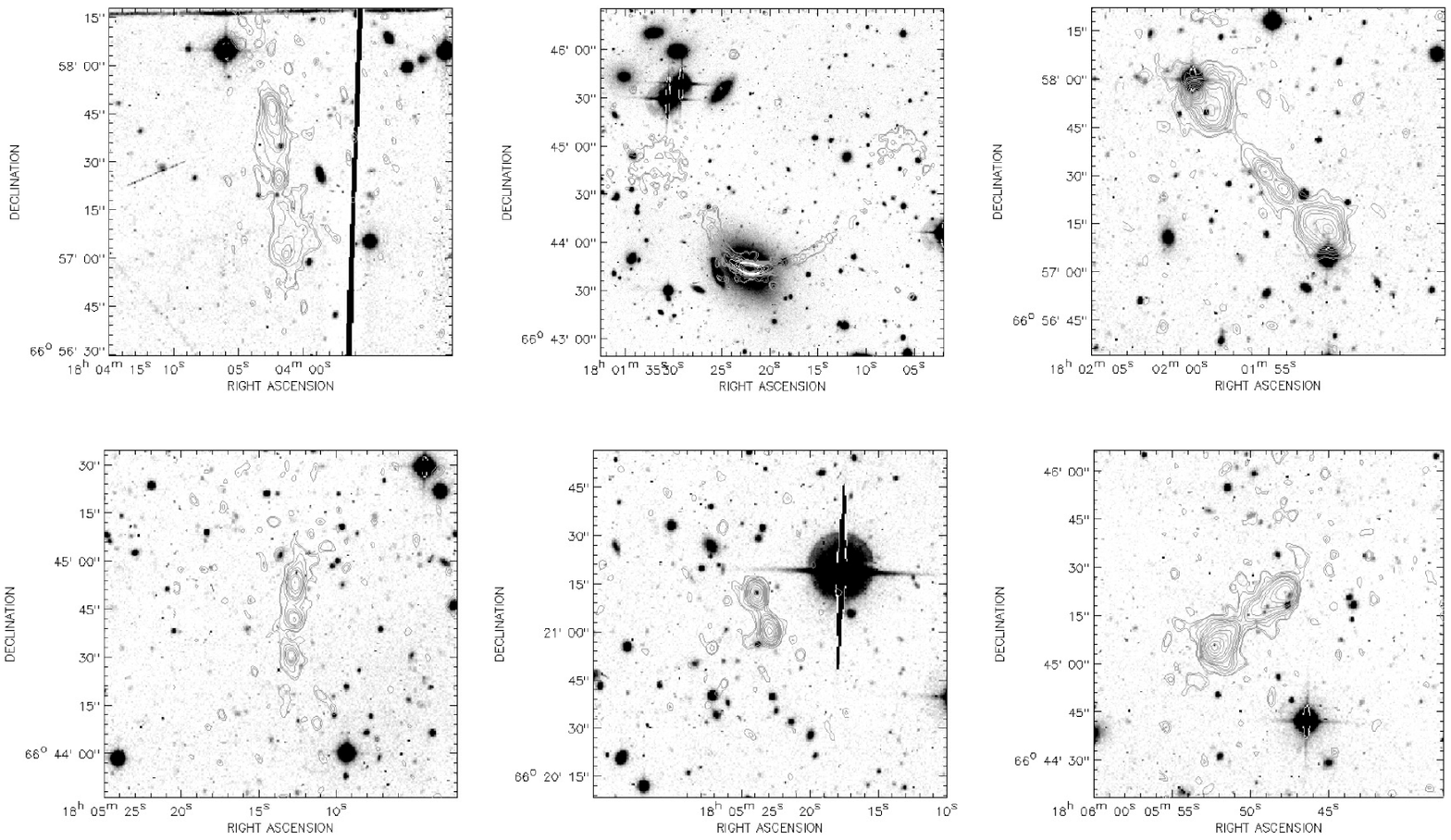}\\
 \caption{Caption the same as for Figure \ref{extendedsources1}.}
 \label{extendedsources2}
\end{figure*}

In the terminology of this paper, a radio component is a region of radio emission represented by a Gaussian-shaped object on the map. Close radio doubles are represented by two Gaussians and are deemed to consist of two components making up a single source. Examples of complex and multi-component radio sources are shown in Figures \ref{extendedsources1} and \ref{extendedsources2}.
Source counts and statistically based samples need to be corrected for multiple component sources so that the fluxes of physically related components are summed together. However, for large-area surveys containing many thousands of radio components, it has often been impractical to visually inspect every component to assess whether these form part of multiple sources.

\begin{figure}
% \centering
\hspace{-1cm}
 \includegraphics[width=1.22\linewidth]{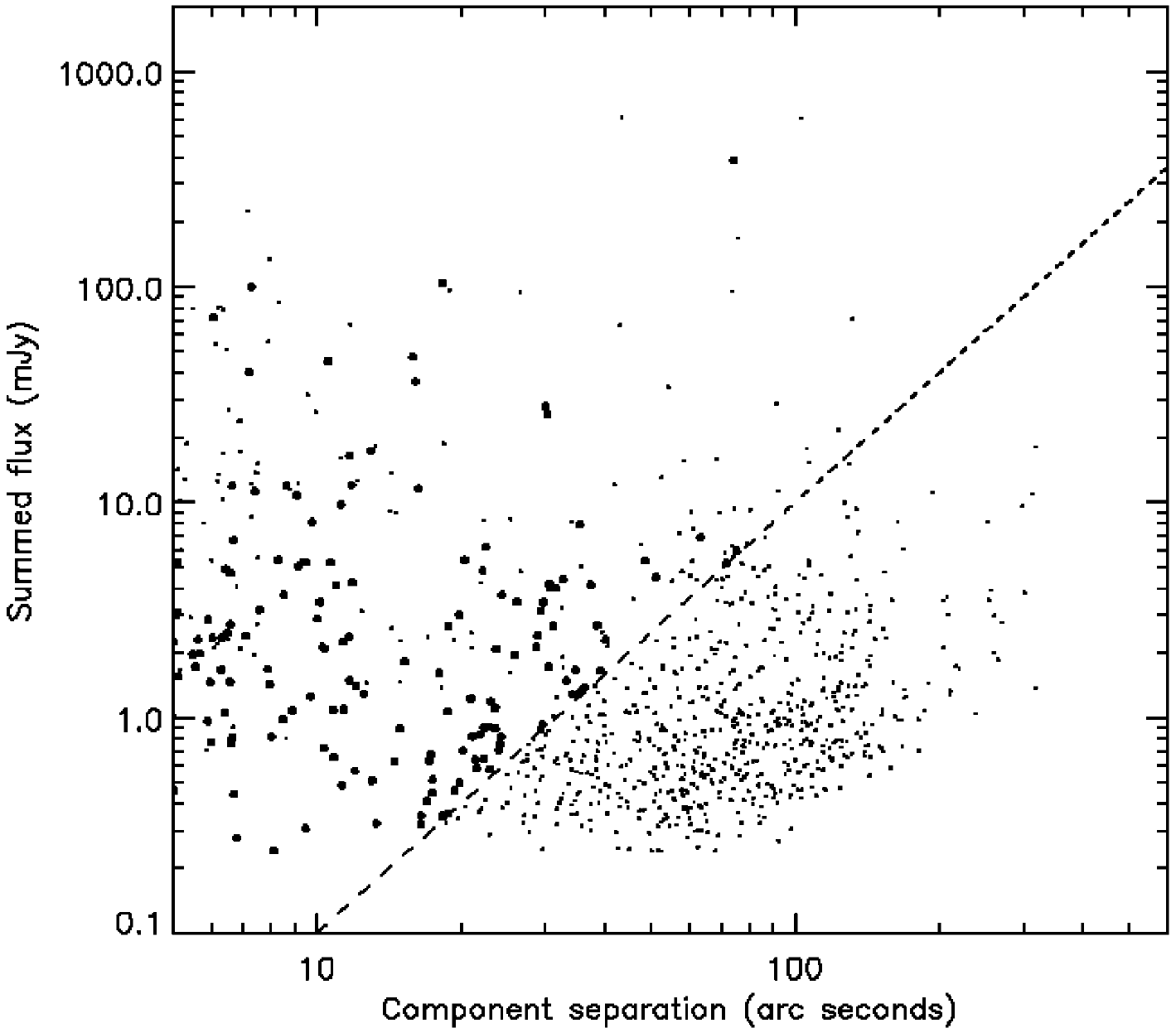}\\
 \caption{This shows the sum of the flux densities of the nearest neighbours between components from the detection catalogue. Near neighbour pairs to the left of the line are considered to be possible double sources. The double sources can be further constrained by requiring the fluxes of the two components $f_1$ and $f_2$ to be in the range 0.25 $\le$ $f_1$ / $f_2$ $\le$ 4, and the 54 radio components, representing 27 radio sources satisfying this criterion are shown as bold circles in the Figure. The smaller dots, mostly in the lower right, points represent single-component sources.}
 \label{nearestneighbour}
\end{figure}

 ~\cite{Magliocchetti1998} have proposed a quantitative approach to identifying the double and compact source populations, plotting the separation of the nearest neighbour of a source against the summed flux of the two sources, and selecting objects where the ratio of their fluxes, $f_1$ and $f_2$ is in the range 0.25 $\le$ $f_1 / f_2$ $\le$ 4. In Figure \ref{nearestneighbour} the sum of the fluxes of nearest neighbours is plotted against their separation.

The dashed line on Figure \ref{nearestneighbour} can be used to further constrain the identification of multi-component sources by setting an absolute flux criterion as defined by \cite{huynh05}:
\begin{equation}
\theta  = 100\left[ {\frac{{S_{\rm tot}~ ({\rm mJy}) }}{{10}}} \right]^{0.5}
\end{equation}
where the separation, $\theta$, is in arc seconds, and the flux is the sum of the components as described above. Using these criteria, 622 radio components in the present survey would be considered to consist of double or multiple sources, according to the criteria from the  \cite{Magliocchetti1998} relationship of the PyBDSF multi-source grouping. However, this is based on clustering properties and the assumption that two adjacent radio components should have flux ratios within an assumed range. To test whether the \cite{Magliocchetti1998} and  \cite{huynh05} criteria do identify true double radio sources, we examined the matches subject to an additional criterion of whether or not the components are associated with a compact optical or an infrared source.

Observations of many extended, or multi-component radio sources often show the presence of optical or infrared emission, where the optical light comes from the associated galaxy or quasar, and the mid-infrared emission from light absorbed by the torus and re-emitted in the mid-IR. The twenty-seven pairs of radio components identified by the \cite{Magliocchetti1998} were then cross-matched with the optical and infrared catalogues. Taking account of the false association rate of 2.6$\%$ discussed in Section \ref{Optical_near_IR}, and the separation between the radio components (typically between 15 and 30 arc seconds), it would be expected that only one (false) identification with the 54 radio components (assuming that the lobes of the radio jets did not emit optical or IR emission would be detectable. The number of cross-matches between the 54 radio components and optical and IR sources was twelve ${\it ALLWISE}$  objects, and seventeen optical objects, of which 8 radio components had both an ${\it ALLWISE}$  and an optical identification.

The same cross-correlation was run centring the position of the radio source at the geometric mean between the two radio components, getting three cross-matches. Based on the \cite{Magliocchetti1998} and  \cite{huynh05} criteria, and the subsequent cross-matching, components identified from the use of these criteria are likely to be double radio sources having an associated central optical or infrared object.

\section{Identifications}

Identifying radio sources with objects visible at other wavelengths is normally achieved by cross-matching within the error ellipse of the radio component (or radio source if it is made up of several components - see Equation \ref{positionaccuracy}), as well as the methodology presented by \citet{Re97}, and developed by \citet{hopkins03} in selecting the optical source with the lowest probability of being an accidental alignment. Studies of optical identifications in the ADF-N such as (\citealt{solarz15}) have indicated that the dominant populations are local star-forming galaxies at a median redshift $\sim$ 0.25, and LIRGs at a median redshift of $\sim$ 0.9 having $L_{IR}~\sim$  10$^{11.62}$~\Lsun. The GMRT 610 MHz data were cross-matched with a compilation of available redshifts for the NEP deep field from existing observations previously mentioned in Table \ref{ancillary}, which are based on direct optical spectroscopic observations, along with photo-$z$ estimated redshifts. Although the detailed classifications are not yet complete, AGNs are identified as either being those detected in X-rays (${\it CHANDRA}$ : \citealt{Krumpe2015}), or by their infrared SEDs with at least one emission line with FWHM $\ge$ 1000 \kmsec ~(or if it is Seyfert 1.5 type, the FWHM of the broad component has FWHM $\ge$1000 \kmsec).  Objects outside of the selection criteria are classified as normal galaxies, or in some cases may be local Galactic objects such as stars, planetary nebulae or HII regions.

An initial cross-correlation between the complete GMRT source component list (as opposed to the biassed 169 source sample mentioned in Section \ref{169Sample}, and the SIMBAD astronomical database with an initial 4\arcsec ~cross-matching radius revealed 286 cross-matches, with eighteen (6.6$\%$) identified as AGN, one BL Lac object, one hundred and six (35.3$\%$) galaxies, four (1.4$\%$) QSOs, one hundred and eleven radio sources (38.8$\%$ with no other identification), one  Seyfert type 2 galaxy (0.35$\%$), three stars (1$\%$ as classified in the spectroscopic survey by \citealt{shim13}), eight X-ray sources (2.5$\%$ with no other cross-identification) and 34 unknown objects (9.4$\%$).

The positional accuracies listed in Table \ref{sourcecatalogueshort} are relative to the self-calibrated and bootstrapped positional reference frame described in Section \ref{section:observations}. Other effects that bias the positions or sizes of sources in radio surveys have already been discussed in Paper 1, to which the reader is referred. An estimate of source dimensions calculated by deconvolving the measured sizes from the synthesised beam is also presented, with Table \ref{sourcecatalogueshort} reporting only those more than double the synthesised beam size. To estimate the systematic uncertainties to the positions, the sources identified as being bright (S$_{\rm{tot}}$ $\ge$ 300 ${\mu}$Jy) single isolated Gaussian components were cross-matched with bright, compact optical or infrared sources. This of course assumes that at least some of the brighter radio sources will be associated with bright compact optical nuclei. The total positional uncertainties of a radio component, after accounting for the systematic errors discussed in Section \ref{positionalaccuracy} are estimated by adding these systematic errors with the statistical positional errors derived from the Gaussian fitting process that is reported in Table \ref{sourcecatalogueshort}.

The strategy for identifying cross-matches with objects identified by other surveys has been to search within a 3$\sigma$ error ellipse, where the size depends on the addition of the sum of the total uncertainties (i.e. statistical and systematic) of each of the surveys. This strategy has the disadvantage that sometimes there will be several sources from other catalogues lying inside the radio error ellipse, which can be addressed by estimating the probability-based association by calculating the formal likelihood function of an association taking into account the magnitude-dependent density of optical or infrared objects close to the positions of radio components (\citealt {ss92, McAlpine2012, Pineau2017, Mingo16}).

\subsection{Optical and near-IR associations}
\label{Optical_near_IR}

\begin{figure}
%\centering
\hspace{-1cm}
 \includegraphics[angle=270,width=1.22\linewidth]{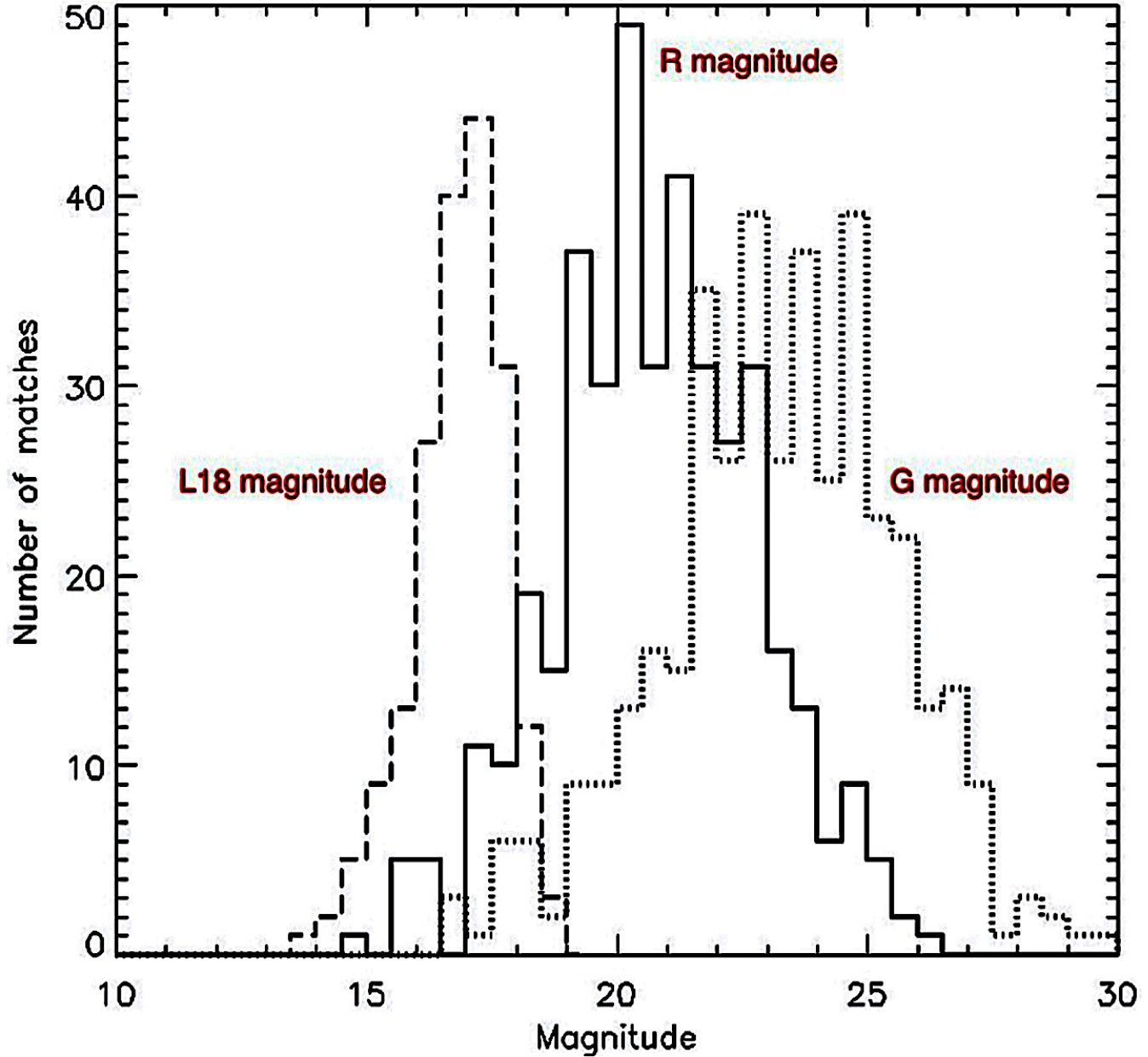}\\
 \caption{The optical and mid-infrared associations with 1286 isolated compact single component radio objects. The vertical axis shows the number of sources in 0.5 magnitude wide bins, and the horizontal axis shows a histogram of the magnitudes of the associated object in the [left to right] 18~$\mu$m (L18) ${\it AKARI}$  filter, and from SUBARU with an $r$- and a $g$-filter, marked R and G respectively on the plot. The peaks of the histograms for the $g$, $r$ and 18~$\mu$m filters are at magnitudes 23.0, 20.7 and 17.2 respectively.}
 \label{Fig:Optmags}
\end{figure}

The radio component catalogue was cross-matched with a composite optical (from SUBARU) and ${\it AKARI}$  near/mid-IR catalogue selected from our ancillary data (see Table \ref{ancillary}), which resulted in 726 cross-matches with an optical source (including the Galactic Catseye Nebula). Of these, 534, 638, and 687 ${\it AKARI}$  cross-matches were found within 1, 2 and 3\arcsec~ respectively of the compact radio components. A similar search was made with the radio positions arbitrarily offset by $\pm$~6\arcmin~in each of the four cardinal directions, resulting in 16 (false) associations, or a false identification rate of 2.6$\%$. Restricting the matches to compact, isolated radio sources, the magnitude distributions of the associated optical object in Figure \ref{Fig:Optmags}.

\subsection{Radio spectral index}

We have compared the GMRT 610 MHz data with the fluxes from our earlier WSRT 1400 MHz survey, and some early LOFAR HBA observations reported in this paper to search for sources showing spectral indices differing from pure synchrotron emission.

\begin{figure}
\centering
 \includegraphics[angle=0,width=1\linewidth]{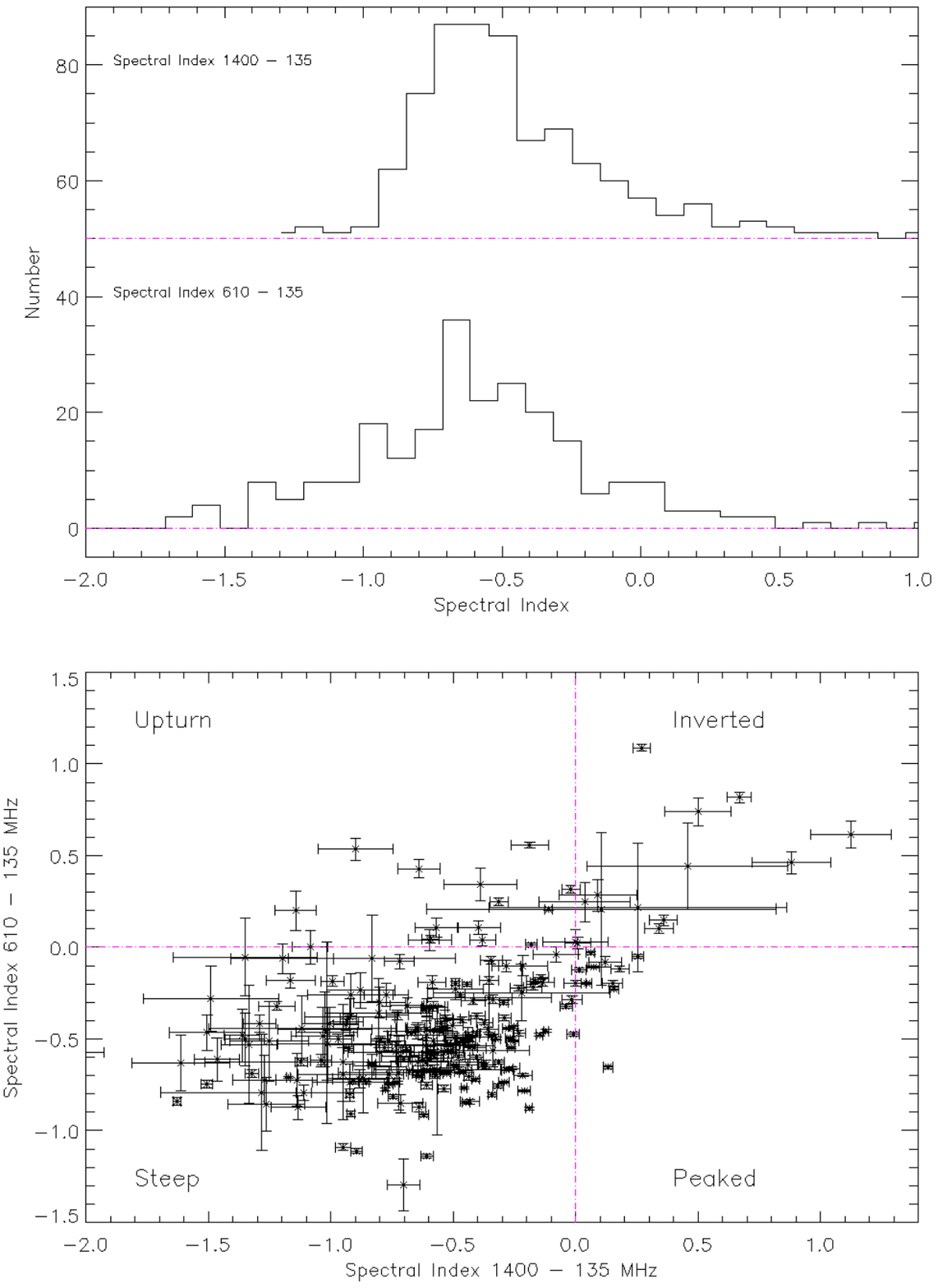}\\
 \caption{The upper plot shows a histogram of the spectral index estimated from 239 compact GMRT sources which were also detected in the WSRT and LOFAR data, with the 1400 - 135 MHz data offset vertically by 50 for clarity.  The lower plot shows a histogram of the two spectral indices, displayed with uncertainties, to identify the spectral shape. }
 \label{specindex_lofar}
\end{figure}

Synchrotron emission is commonly described by a power-law function $S_{\nu}$~$\sim$~$\nu^{~\alpha}$, where $S_{\nu}$~is the radio flux density at a given frequency, $\nu$, and $\alpha$ is the corresponding power-law slope. Therefore the spectral index between, for example, the 610 MHz GMRT and the 135 MHz LOFAR data is given by:
\begin{equation}
\alpha = \frac{ln\left(\frac{S_{\scriptsize 610}}{S_{\scriptsize 135}}\right)}{ln\left(\frac{\nu_{\scriptsize 135}}{\nu_{\scriptsize 610}}\right)}
 \label{specindex1}
\end{equation}
and the associated error on the spectral index is given by:
\begin{equation}
\delta\alpha= \frac{1}{ln\left(\frac{\scriptsize  610}{\scriptsize 135}\right)}\sqrt{~\left(\frac{\delta S_{\scriptsize 610}}{S_{\scriptsize 610}}\right)^2+\left(\frac{\delta S_{\scriptsize 135}}{S_{\scriptsize 135}}\right)^2}
 \label{specindex2}
\end{equation}
where $S_{135}$ and $S_{610}$ are the LOFAR and GMRT fluxes and errors respectively. The 135 - 1400 MHz spectral index can be calculated in a similar way by modifying Equations \ref{specindex1} and \ref{specindex2}, and is shown in Figure \ref{specindex_lofar}.

\begin{figure}
%\centering
\hspace{-1.4cm}
 \includegraphics[angle=0,width=1.3\linewidth]{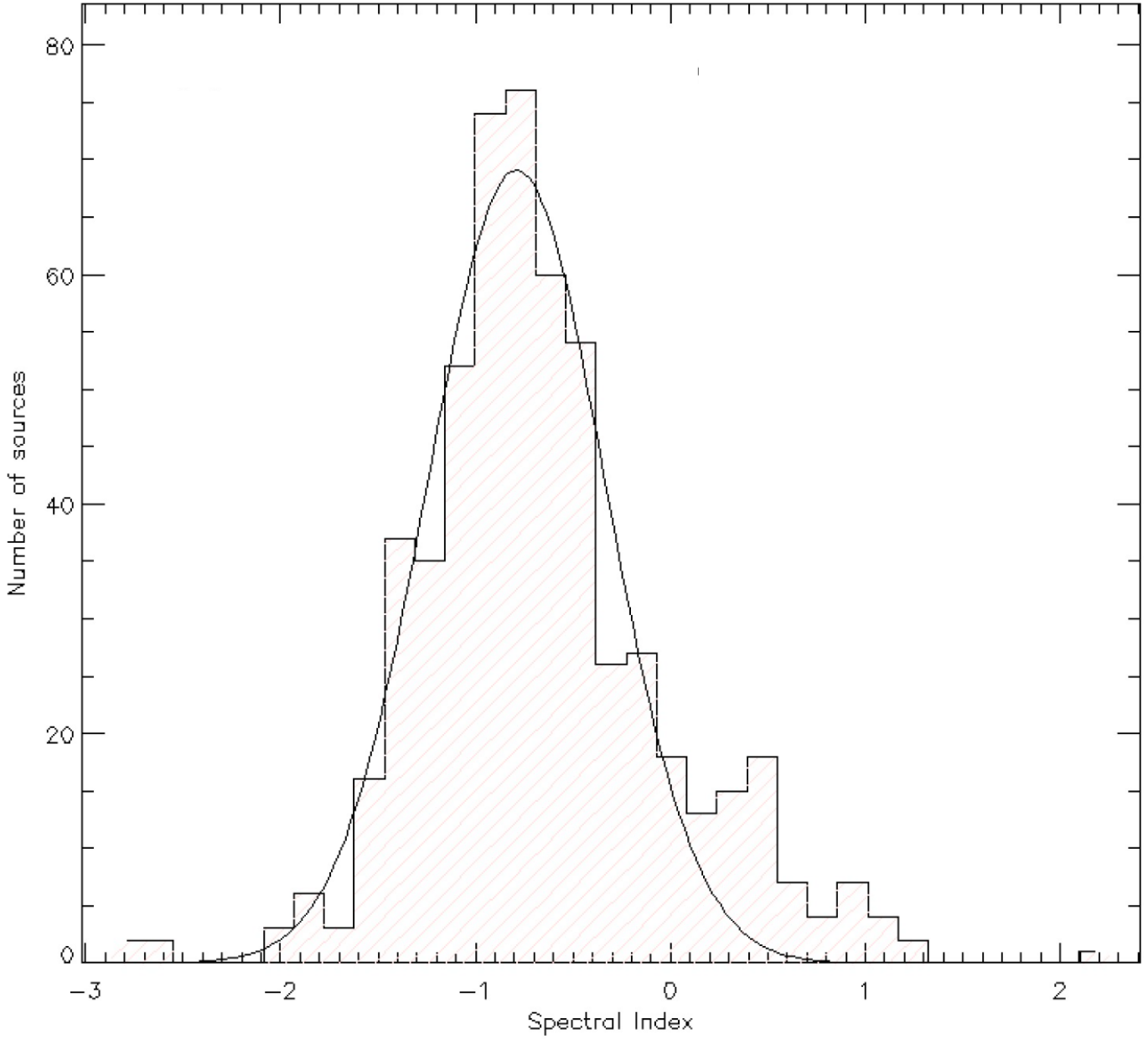}\\
 \caption{Spectral index from 562 cross-matched sources from the GMRT 610 MHz and WSRT 1400 MHz surveys, which have a mean spectral index of -0.79, and a half-width of 0.45 from the superimposed Gaussian fit. Sources with a spectral index $>$ 0 indicate evidence of a spectra turndown at lower frequencies. Whilst the synchrotron emission in star-forming galaxies typically shows a slope of $\alpha$~=~-0.7 \citep{yun01,condon02}, changes in the CR energy distributions or the environmental or ISM processes (such as free-free absorption, ionisation losses, and synchrotron self-absorption) are normally cited as defining the spectral shape at low frequencies.}
 \label{specindex}
\end{figure}

The spectral index formed from the WSRT and GMRT data sets is shown in Figure \ref{specindex}, which shows a small excess of flat-spectrum components with spectral indices $>$ 0. The Catseye Nebula (a Galactic Planetary Nebula close to the field centre) is one such case of a flat spectrum object, but the others, believed to all be extragalactic, indicate that up to $\sim$20$\%$ of the total number of radio components show some evidence for spectral flattening toward lower frequencies.

Figure \ref{specindex_lofar} displays a more detailed look at a small sample of 239 compact sources in the GMRT, WSRT and LOFAR data, to distinguish between steep, peaked, upturn and inverted spectra.

As expected, most sources are located in the area labelled Steep, which is typical of a power-law synchrotron spectrum. A relatively small number however show mildly inverted, upturning or peaked spectra, resembling results from the Lockman Deep field discussed by \cite{Mahony2016}. Of those sources falling within the Inverted classification, source number 680 in Table \ref{sourcecatalogueshort} has a spectral index 1400-135 = 0.66 and spectral index 610 - 135 = 0.82 and is a known QSO. Therefore it is likely that the inverted spectral index may be a consequence of variability rather than of the emission mechanism. This applies to all of the objects where the spectral index estimated from surveys taken at different epochs may be unreliable as an indicator of the emission mechanism if the source intensity varies between the data sets, as is likely in the case of the associated QSO.

\subsection{Redshifts}
\label{Redshifts}

Many different methods to calculate photometric redshifts by fitting to the spectral energy distribution (SED) have been adopted for situations where spectroscopic redshifts are not available. The uncertainties in this approach as mentioned previously, affect almost every other research contribution using a similar approach. To explore the efficacy of various approaches, several different SED fitting codes were tested against each other and against a sample of sources from the GMRT catalogue for which spectroscopic redshifts are also available (\citealt{Oi2014}). For the present paper the photometric redshift code  CIGALE\footnote{https://github.com/JohannesBuchner/cigale} \citep{No09}, was used to estimate the physical parameters of the radio-source host galaxies, such as AGN fraction, with the latter being estimated from the flux across the 0.1-5 micron region from the original definition by \citet{Fritz06}.

The photometric redshift catalogue available in the NEP field (using the LePhare photometric redshift estimator which only uses optical data: \citealt{Oi2014}) can be seen to be a good proxy for two cases a) for z~$<$~1 if the magnitude in the CFHT $z$-band is $z_{mag}$$<$~24, and b) z~$<$~2.2 if the magnitude $z_{mag}$~$<$~22. However, we note that the accuracy of photo-z's depends on the source magnitudes (and redshifts), as the photo-z of a bright object can be accurately estimated at redshifts up to z $\sim$ 2 and photo-z of a faint object is reasonable at redshifts z $<$ 1. This may bias the results since there are likely to be at least some faint radio-source hosts and high-z sources in the GMRT sample, and this catalogue does not guarantee a reliable redshift in the radio sample. In addition, the LePhare code focuses on the use of templates of nearby galaxies, which may not necessarily be representative of the wider population of high-z sources.

\section{Analysis of a 169 source sample}
\label{169Sample}

In this paper, a multi-wavelength catalogue of 169 radio sources (identified in Table \ref{sourcecatalogueshort} with a star symbol (*) added to the running number) with available redshifts from our ancillary data was used to test and train the CIGALE-based algorithm. For a full discussion of the various inputs and outputs from CIGALE, the reader is referred to the original descriptor paper by This sample is a subset of the GMRT catalogue which includes optical-NIR data from the CFHT, IR data from ${\it AKARI}$  and sub-millimetre wavelength data from  ${\it HERSCHEL}$ -SPIRE. Ninety per cent of the sources had at least 10 photometric detections. This sample was based on the availability of cross-matching data, and should not be assumed to represent a statistically selected sample based on intrinsic physical properties or sensitivity limits. The inherent problems concerning almost all research contributions where combined optical/NIR and FIR/submillimetre (sub-mm) photometry data points are used in SED fitting are well reported in many studies: in particular, the different angular resolutions of these various datasets, meaning that the FIR/sub-mm photometry may include more diffuse/unresolved background emission, or that some of the sub-mm sources may have more than the assumed single identified counterpart or that other effects such as gravitational lensing may modify the apparent fluxes at different wavelengths. With those caveats, we examine the properties of the 169 source sample.
\subsection{CIGALE estimation of the 169 source sample}
\label{CIGALE169}

\begin{figure}
 \centering
 \hspace{-0.3cm}
 \includegraphics[angle=-90,width=1.04\linewidth]{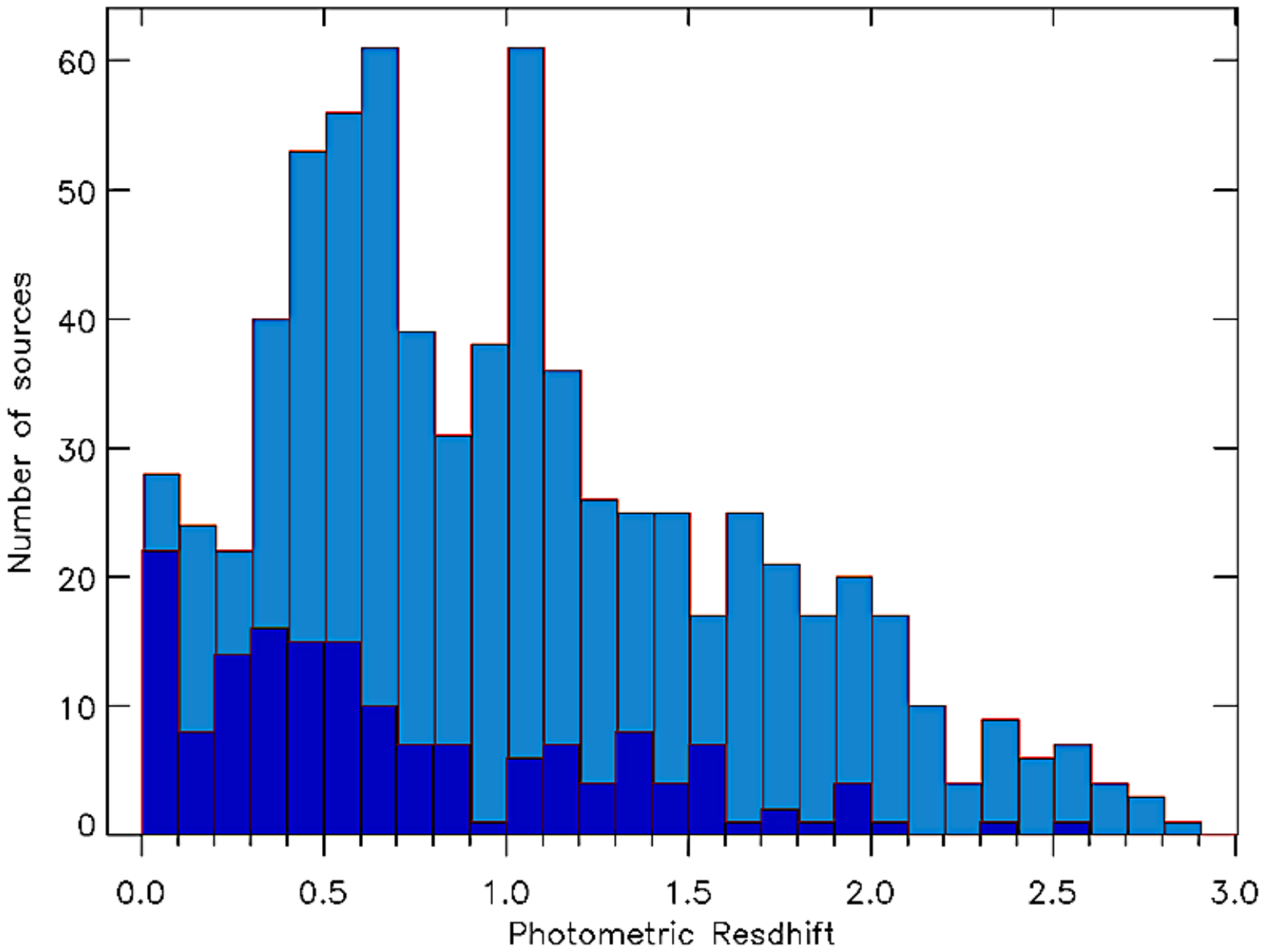}\\
 \caption{The redshift distribution of 726 cross-matched radio sources with their estimated photometric redshifts using CIGALE shown in light blue and overlaid with the redshift distribution of the 169 source sample discussed in Section \ref{169Sample}, shown in dark blue. As discussed in the text and shown in Figure \ref{CigaleLePhare}, the photometric redshifts become less reliable compared to spectroscopically confirmed redshifts of about 1.2.}
\label{PhotoZ}
\end{figure}

CIGALE was used to estimate the photometric redshift by calculating a grid fitting the available data from 0.1 $<$ z $<$ 5 with steps $\delta$$z$ = 0.1. This technique results in the redshift distribution of the 726 cross-matched sources as shown in Figure \ref{PhotoZ}.

\begin{figure}
 \centering
 %\vspace*{-3.3 cm}
 \includegraphics[angle=270,width=1\linewidth]{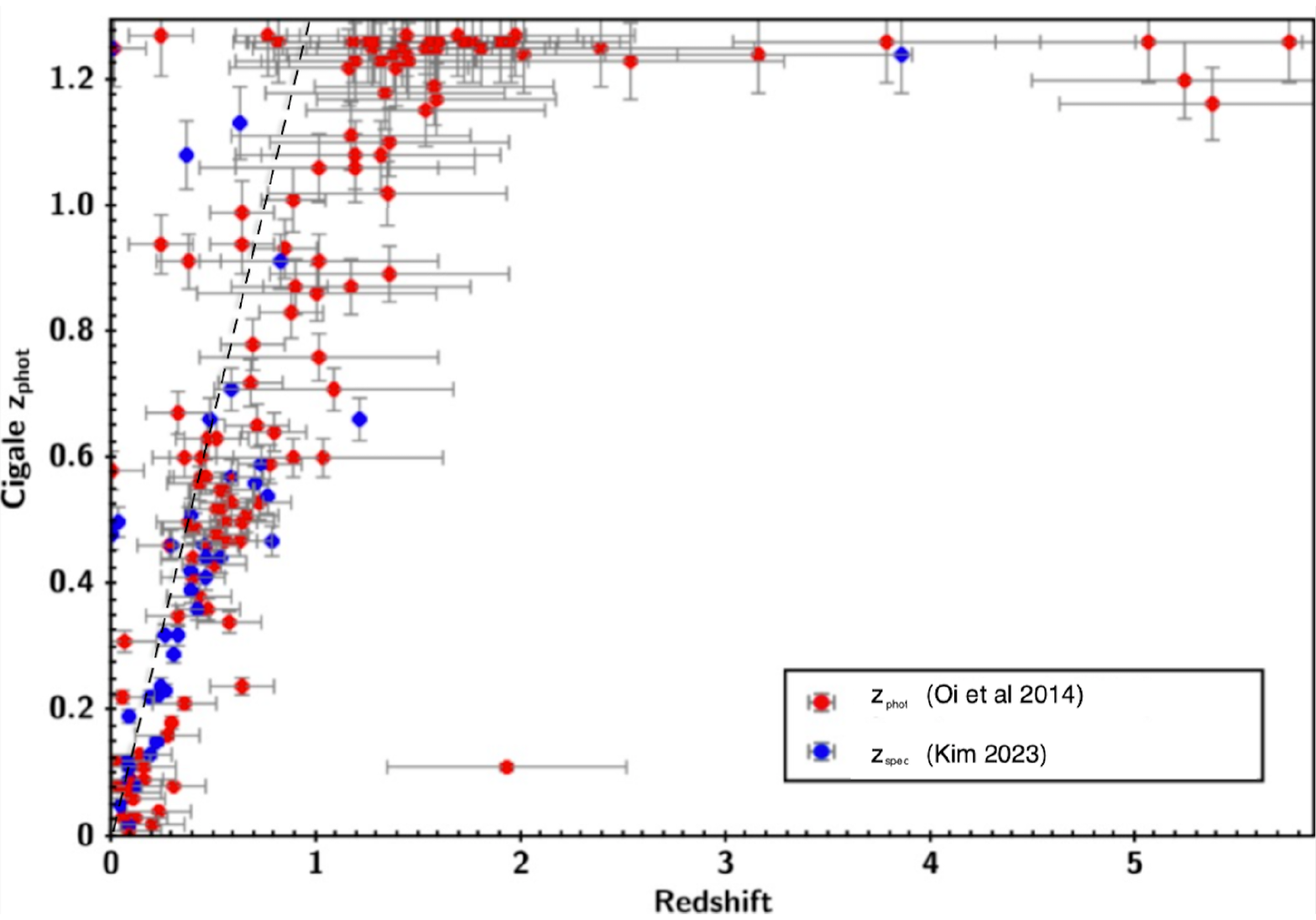}\\
 \caption{This plot compares photometric redshifts estimated from CIGALE, with the direct spectroscopic redshifts shown as blue squares and photometric redshifts in red circles \citet{Oi2014} and the spectroscopic observations \citet{Kim23}. There is some agreement as far as z~$\sim$~0.7 (see the 1:1 dashed line), but the two estimates are known to deviate at higher redshifts \citet{Oi2014}.}
\label{CigaleLePhare}
\end{figure}

For the spectroscopic redshifts to be reliable, at least two emission lines have to be covered by the photometric data, otherwise, there is the possibility for systematic errors to bias the other parameters. The photometric redshift calculated by using optical data alone with LePhare was not found to be sufficiently robust, and it was decided to leave the redshift as a free parameter. The CIGALE photometric redshift estimates were compared with direct spectroscopic redshifts from \citep{Oi2014}, and with LePhare for faint and high-z sources, in conjunction with a preliminary spectroscopic redshift catalogue in the NEP \citep{takagi10}. This showed that both sets of photometric redshifts are in reasonable agreement up to $z_{phot}$ $\sim$ 0.7 in the case of the \citep{Oi2014} estimates that were based on high-quality spectroscopic redshift measurements as shown in Figure \ref{CigaleLePhare}.

\begin{figure*}
 \centering
\hspace*{-2.5 cm}
 \includegraphics[angle=-90,width=1.25\linewidth]{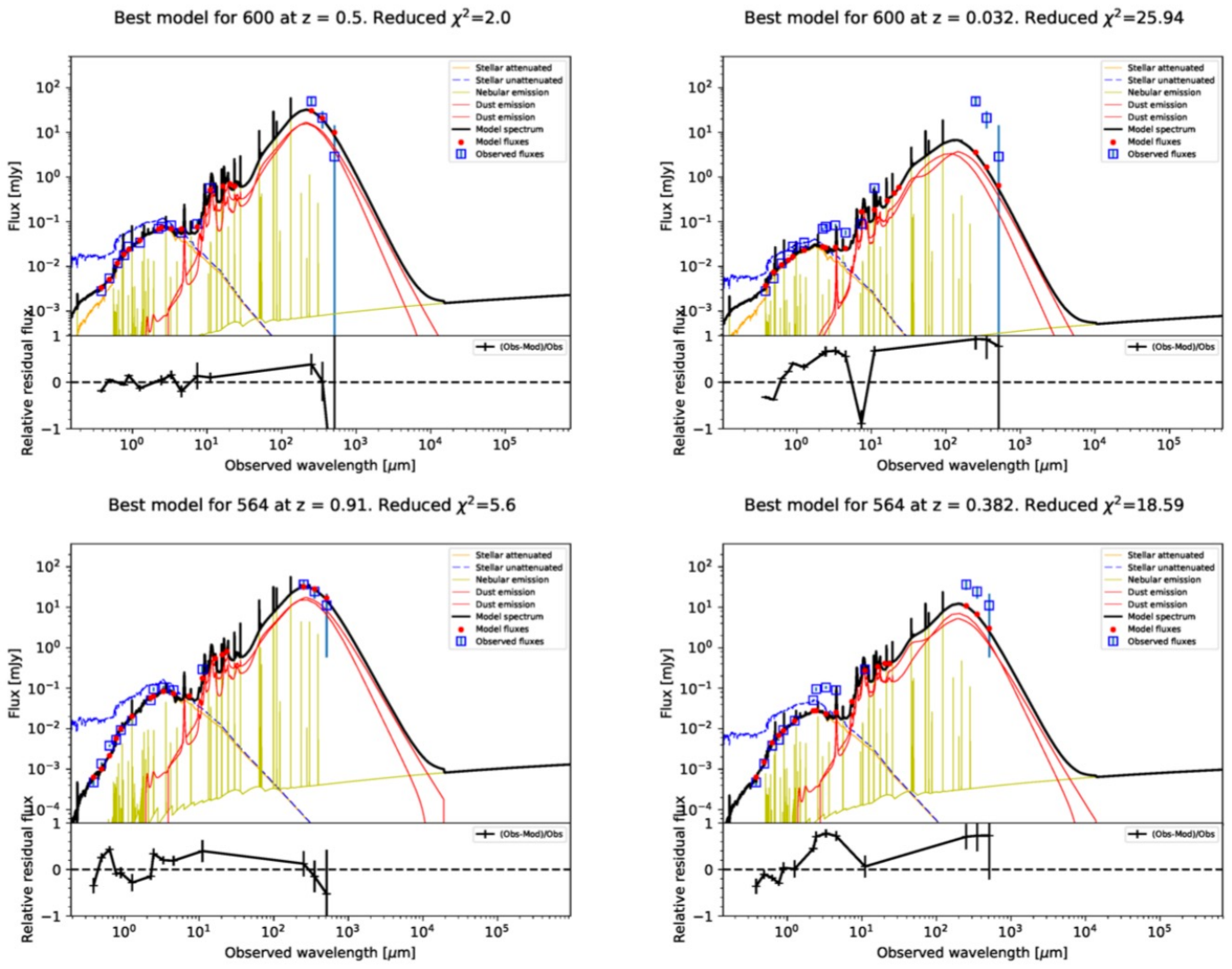}\\
 \caption{This shows the SED from the optical - sub-mm wavelengths of radio source associations in each row. The first column of the upper row shows the SED estimated using the photometric redshift calculated with CIGALE, whereas in the second column, the SEDs calculated with the redshift data available in the field are shown. The upper plot row shows the SED fitted with the photometric redshift ($z_{phot}$ = 0.5) obtained with CIGALE (on the left) and the same source fluxes fitted with the spectroscopic redshift ($z$ = 0.032) \citet{takagi10} as an input which has a redshift flag: D = probable redshift but unreliable. The results are better if the redshift was estimated with CIGALE than with an uncertain (flag D) spectroscopic redshift: the FIR observations are clearly in better agreement in the top SED and the $\chi^2$ = 2.0 vs. $\chi^2$ = 25.94 show better fit (both have the same photometric detections, 15 in total). The lower plot row shows a comparison of the photometric redshift calculated with CIGALE as a free parameter ($z_{phot}$ = 0.91) with the SED from the photometric redshift catalogue of \citet{Oi2014} which has $z_{phot}$ = 0.382, where the $\chi^2$ is three times higher and the model fits neither the FIR peak nor the optical peak (both have 13 photometric detections). The two different dust models shown are the [upper in the plot legend] \citet{dl07} and \citet{schr16} models. The former takes into account the emission from small dust grains and the characteristics of intense MIR emission, whereas the latter model is a simpler model of the PAH emission, and is more focused on the FIR-submillimetre peak, allowing higher dust temperatures. The source Best model number refers to the catalogue running numbers indicated with a star in Table \ref{sourcecatalogueshort}.}
\label{CigaleSED}
\end{figure*}

SEDs were also calculated using CIGALE by giving the redshift from \citet{Oi2021} as an input. The code was able to find a model that fits 77$\%$ of the sources, however, when the redshift is a free parameter, the code found models for all of the sources in the sample with better SED fitting. SED fitting was made for both catalogues along with the photometric redshift estimated from CIGALE and compared by eye for the 169 source sample. For all the sources a better (or equal) quality fit was produced using the CIGALE photo-z input as shown in Figure \ref{CigaleSED}.

CIGALE allows us to model the SED extending from UV to radio wavelengths by balancing the energy absorbed by dust and then remitted at IR wavelengths. A more detailed explanation of the code and module selection is described in by \citet{No09}. This code assumes a combination of modules that model different physical aspects of the spectra, according to the approximations shown in Table \ref{CigaleSEDinputs}.

\begin{table*}
\centering
\begin{tabular}{lll}
\hline
\textbf{CIGALE module} & \textbf{Main parameters} & \textbf{Description} \\ \hline
SFH                    & $\tau_{main} = 250 - 8000$   & Main stellar population [Myr] \\
Stellar emission       & IMF=1, $Z=0.02$              & Chabrier IMF, Metallicity \\
Dust attenuation       & $A_{Vyoung} = 0 - 4$,       & Attenuation V band \\
                       & powerlaw$_{\rm slope}$ = -0.7,        & Slope delta of the power law \\
                       & $A_{Vold} = 0.1 - 1$,       & Reduction factor compared to $A_{Vyoung}$ \\
                       & $UV_{bump\lambda} = 217.5$          & Central $\lambda [nm]$ of the UV bump \\
Dust emission models   & $q_{PAH} = 0.47 - 5.26$,     & Mass fraction of PAHs \\
                       & $U_{min} = 1.05$           & Minimum radiation field \\
AGN component          & fracAGN = 0 - 0.5,           & AGN fraction \\
                       & $r_{ratio} = 60 - 140$,       & Opening angle of the dust torus \\
                       & psy = $0^\circ - 90^\circ$   & Angle between AGN axis and line of sight \\ \hline
\end{tabular}
\caption{This shows a list of the input parameters used in CIGALE for the analysis of the 169 radio-source sample. The star formation history (SFH) assumed is a decreasing exponential. The stellar emission model is commonly used and defined by the Chabrier initial function mass \citep{bc03}, with metallicity similar to the solar (Z$_{\sun}$). The dust attenuation is defined by a power law (\citealt{cf00}) that allows for a wide range of PAH values. The AGN component assumes a broad range of parameters such as the AGN fraction and physical parameters of the torus \citep{Fritz06}. The CIGALE modelling used in this work varies all of the Main parameters listed above to achieve best fits.}
\label{CigaleSEDinputs}
\end{table*}

The star formation history (SFH) was modelled as a declining exponential star formation rate SFR, and so is not expected to have high star formation and it is not necessary to introduce the presence of a starburst to fit the SED. The stellar emission contribution was chosen following \citet{bc03}, and setting the Chabrier initial mass function and metallicity similar to Z$_{\sun}$. The dust attenuation assumes a power law (\citealt{cf00}), and includes a correction for the 'blue bump', which is related to the accretion disk at the centre of AGN. The \citet{dl07} models were adopted to define the dust emission, introducing the presence of PAHs and the radiation field. Since one of the aims was to study the mean AGN contribution in radio galaxies, the AGN emission was specifically defined with broad grids in several parameters and varying small steps in the values. The nebular emission - which models the emission lines - was considered in the first test where the photometric redshift was considered but was not found to contribute significantly to the main parameter estimates. For this reason, it was removed from the rest of the analysis since it was computationally expensive, increasing the models produced by several million. In total, this combination of physical parameters resulted in 33 million possible models, from which we chose the best fit for each source.

\subsection{Source properties estimated from CIGALE analysis}

The sub-mJy radio sources are most commonly associated with star formation since the radio emission is believed to be dominated by the emission from supernovae events. However there is a mix of galaxy populations around these flux levels that are not well characterised, and the evolutionary dependence of star-forming galaxies and AGN remains poorly constrained \citep{padovani09}. Also, the 169 source sample was chosen as having either photometric or spectroscopic redshift information of a presumed optical host.

\begin{figure}
 \centering
\hspace*{-0 cm}
 \vspace{-0 cm}
 \includegraphics[angle=0,width=1.0\linewidth]{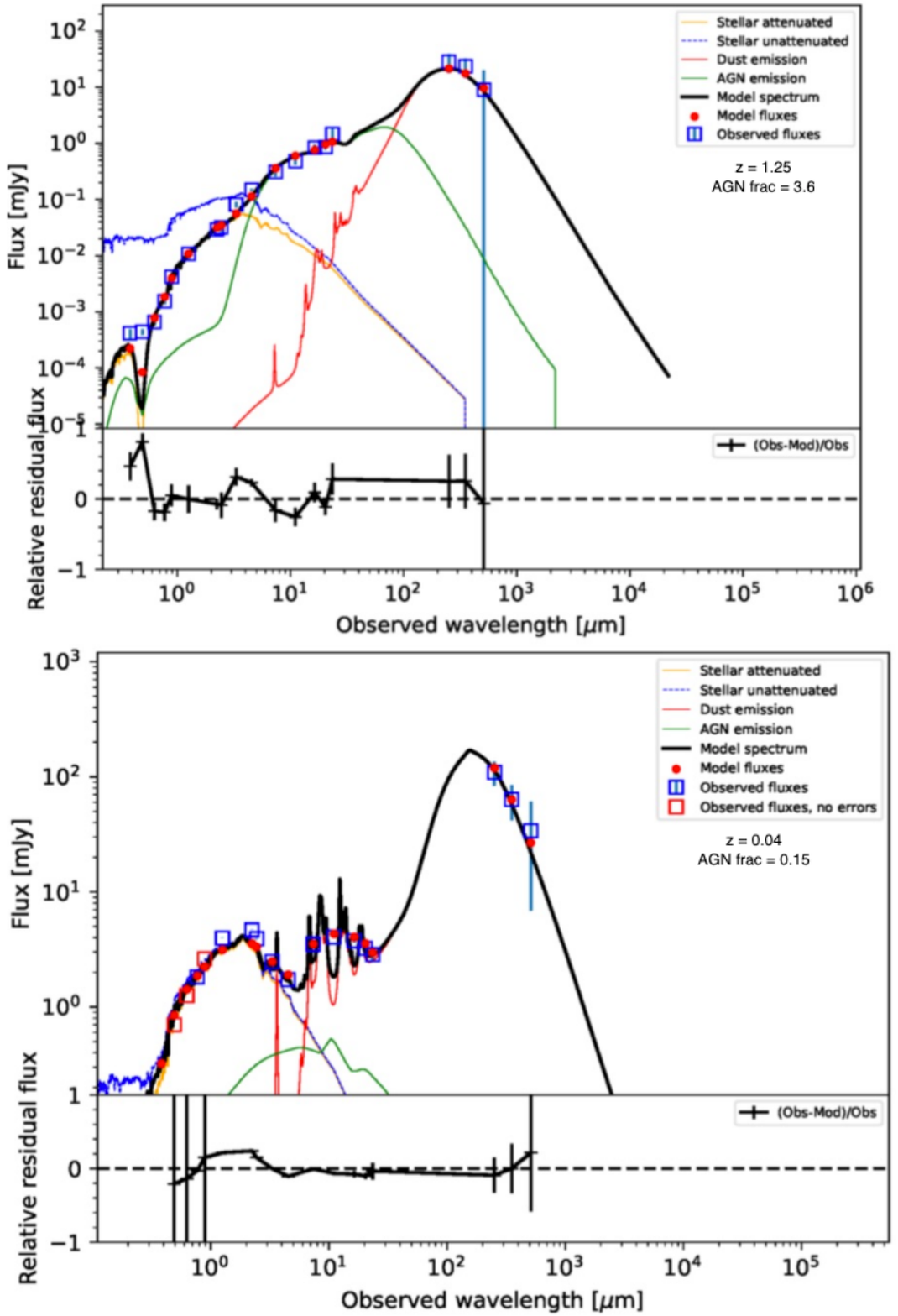}\\
 \caption{Examples of SED fitting to the optical - sub-mm data for sources associated with GMRT radio sources, as estimated with CIGALE. The blue squares are the observed fluxes [in mJy units] and the red dots show the predicted fluxes by the best model that fits the observations (black line). CIGALE modules are listed in the legend: the blue line is the stellar emission (\citealt{bc03}), the yellow line defines the dust attenuated stellar component, whereas the dust emission is shown by the red line from \citet{dl07}, and the green line defines the AGN emission (\citealt{Fritz06}). [Top] This plot shows the highest redshift source found $z_{phot}$ = 1.25), corresponding to GMRT source 454 in the running number of the extended Table \ref{sourcecatalogueshort}. It clearly shows a power-law spectrum, which together with the AGN fraction (50$\%$) indicates that the galaxy hosts an AGN. The fit is robust ($\chi^2$ = 6.47 for the 18 photometric detections), and there is an agreement between the optical/near-IR and the FIR peak, which indicates that the cross-match is consistent. [Bottom] Example of a representative source of the sample at low redshift ($z_{phot}$ = 0.04), corresponding to GMRT source 101 in the running number of the extended Table \ref{sourcecatalogueshort}, showing the presence of PAH emission and a lower AGN contribution (green line).}
\label{CigaleGood}
\end{figure}

The Bayesian analysis using CIGALE gives the most probable values for all of the relevant parameters described by the models. To study the sample of galaxies with faint radio emissions the following parameters were considered for the probabilistic analysis: star formation rate (SFR), stellar mass, gas mass, AGN fraction, PAH mass fraction and luminosity of the dust as being the most important. This study is focused on estimating the physical properties of galaxies associated with faint radio emission, especially the AGN contribution and star formation rates in these sources. A full description of the various inputs and outputs from CIGALE, and the definitions therein of other quantities, such as for the SFR, is given by \citet{boquien19}. Examples of good quality fits are shown in Figure \ref{CigaleGood}.

The SEDs of the 169 source sample were checked by eye to ensure good agreement across the whole spectrum. Another approach might have been to use only the optical-to-NIR photometry to fit the SED, which should result in a more reasonable redshift, stellar mass, and SFR, or to adopt some de-blending methods (e.g., the de-blending method developed in \citealt{jin18, liu18}) to ‘de-blend’ FIR to sub-mm photometry at the positions from high-resolution optical/NIR surveys. Since the main results of this work are focused on the radio catalogue and some exemplar SED analysis based on best-fit SEDs of sub-mm sources, it was felt that a more detailed analysis was not warranted at present. Although models were found for all of the sources, three of them were excluded since the estimated SEDs that did not fit the whole spectrum provided an uncertain mismatch between the optical and the FIR data.

The photometric redshift distribution of the radio sources extends over the range 0.1 $<$ $z_{phot}<$ 1.25, with a mean $z_{phot}$ = 0.55 $\pm$ 0.03, and the highest concentration of sources at $z_{phot}$ $\sim$ 0.1 (shown previously on Figure \ref{PhotoZ}). The redshift distribution of the 169 source sample has few sources at high redshift, which could be due to the sub-sample of sources being biased by having either a spectroscopic or a photometric redshift, and therefore may not be statistically representative of the intrinsic radio population \citep{Oi2014} (see Figure \ref{CigaleLePhare}).

\begin{figure}
 \centering
  \includegraphics[angle=-90,width=1\linewidth]{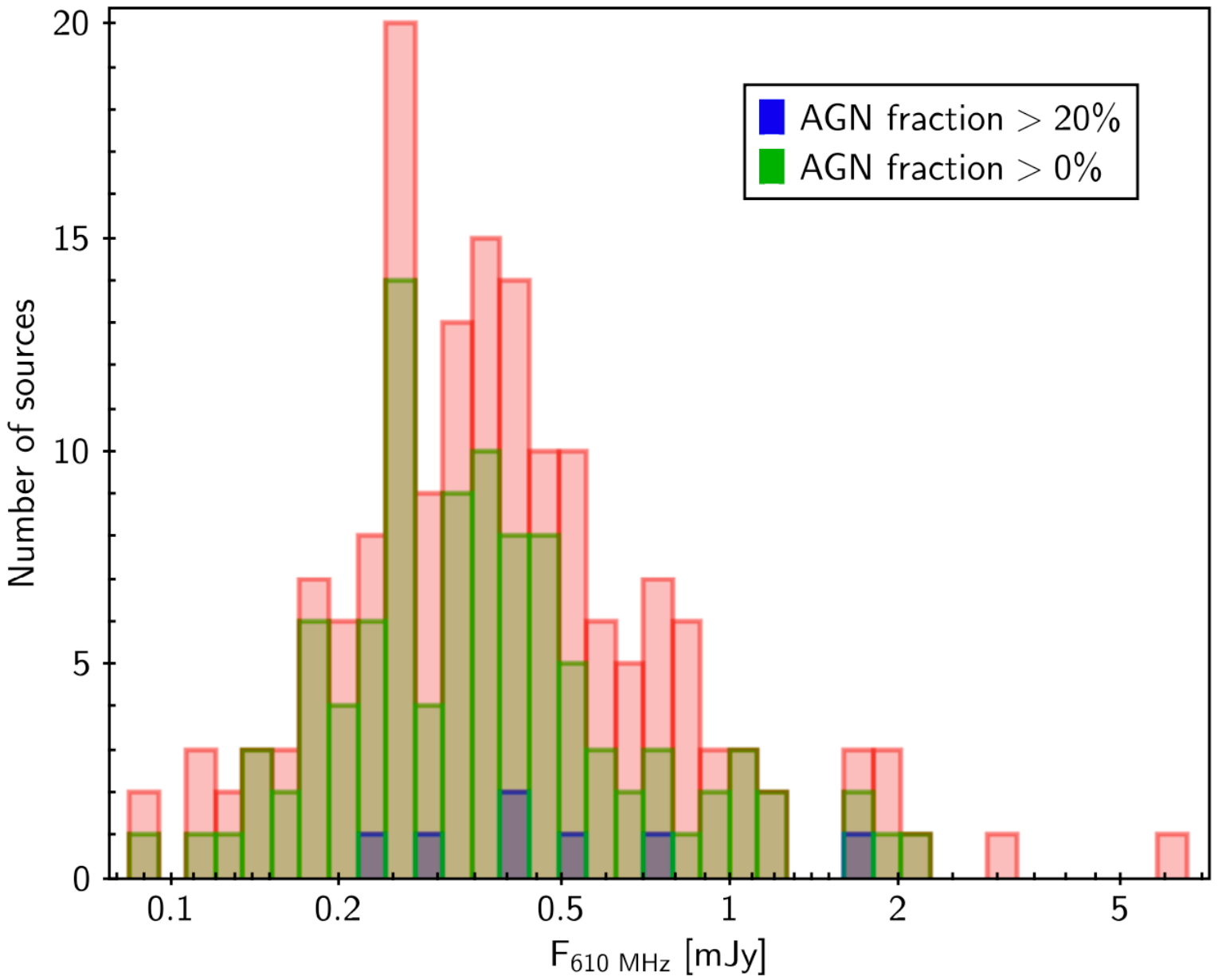}\\
 \caption{The observed radio flux histogram shows the distribution of faint radio sources, 92$\%$ of which have $S_{610}$ $<$ 1mJy. The total number of radio sources is plotted in red, whereas the sources with AGN contribution are over-plotted in green (AGN fraction $>$ 0$\%$ as defined by \citealt{buat15}) and in blue (AGN fraction $>$ 20$\%$).}
\label{radiofluxhistogram}
\end{figure}

Amongst the 169 source sample, 92$\%$ have $S_{610}$ $<$ 1~mJy, and therefore the sample is more likely to trace the sub-mJy population of star-forming dominated objects than AGN-powered sources \citep{padovani09}. In the literature (see \citealt{windhorst85} for review) the micro-Jy population is made up of a mix of galaxy types: starburst, post-starburst and elliptical galaxies. The radio emission at these fluxes mainly results from the presence of (nuclear) starbursts and weak AGN activity. Possibly due to this fact, the distribution of the radio sources from the 169 source sample shows no clear trend with AGN fraction (see Figure \ref{radiofluxhistogram}), and a larger sample of sources would be needed to draw firm conclusions.

\begin{figure}
 \centering
 \includegraphics[angle=0,width=1\linewidth]{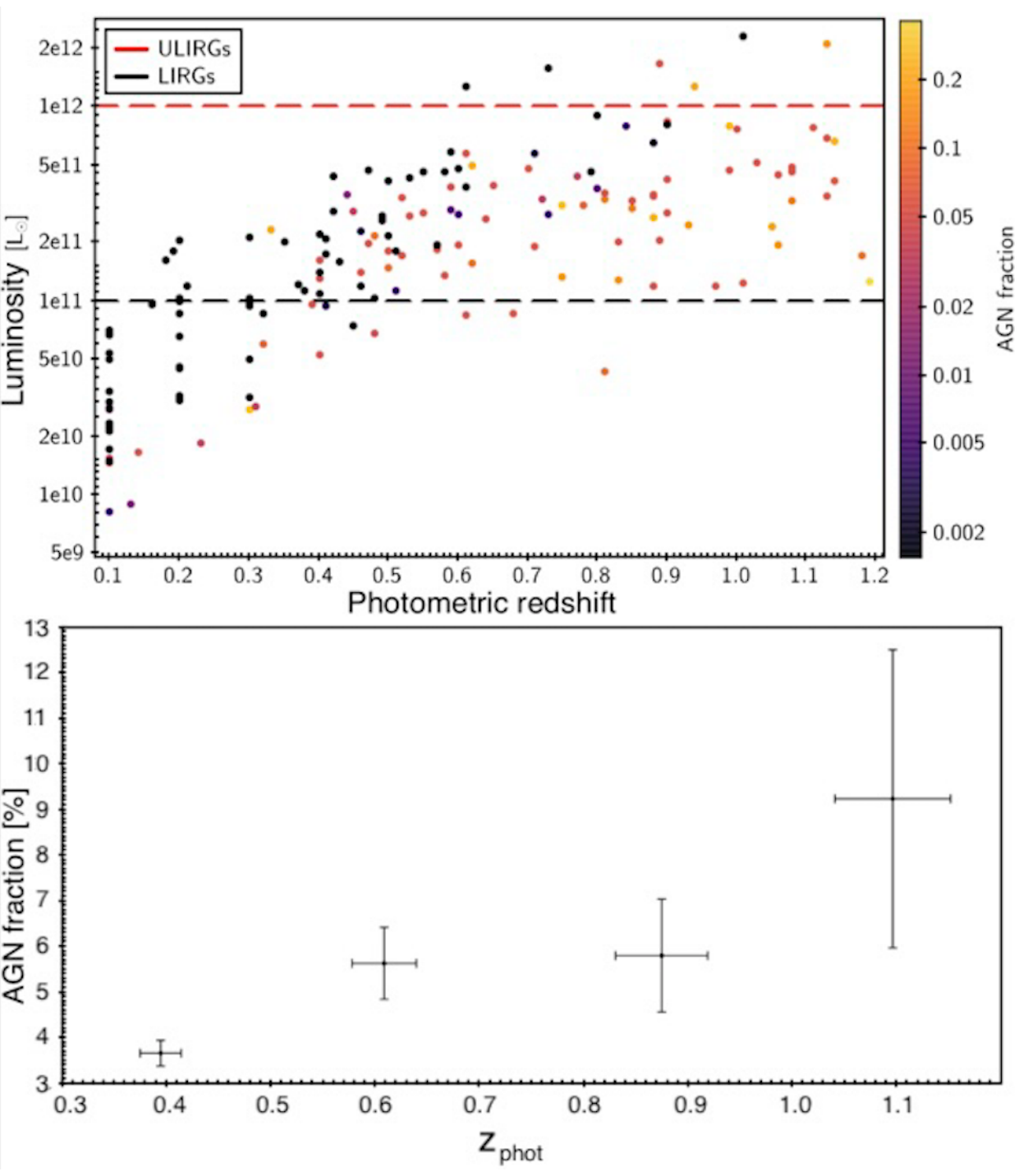}\\
 \caption{[Top] Infrared luminosity - corrected for AGN contribution against the photometric redshift. The sample is divided by luminosity in the plot, where the black dashed line shows the limit for LIRGs and the red line shows the limit for ULIRGs (L$_{IR}$ $>$ 10$^{12}$  \Lsun). The AGN fraction is given in the third axis; it has a trend with luminosity and redshift: higher AGN fractions correspond to higher redshifts and luminosities. The photometric errors are between 0.005 $<$ e$_{zphot}$ $<$ 0.06, whereas the uncertainty in the IR luminosity corresponds to an average of   $\sim$~11$\%$. [Bottom] Mean AGN fraction against redshift (with four redshift bins) for the LIRGs subsample.}
\label{LFIR}
\end{figure}

The far-IR luminosity (8 - 1000 $\mu$m) is proportional to the star formation rate. The different ranges of these IR luminosities are classified with infrared luminosities L$_{IR}$ as follows: LIRGs in the range L$_{IR}$ = 10$^{11}$ - 10$^{12}$ \Lsun), ULIRGs (L$_{IR}$  = 10$^{12}$ - 10$^{13}$ \Lsun)~and hyper-luminous IR galaxies (HLIRGs) (L$_{IR}$ $>$ 10$^{13}$ \Lsun) (see \citealt{lon06} for a detailed review). The radio sample was classified by infrared luminosity corrected for AGN contamination to study the characteristics of these populations (see Figure \ref{LFIR}).

The classification of the 169 radio sample is 4$\%$ ULIRGs, 66$\%$ LIRGs and 24$\%$ of lower luminosity and 6$\%$  of indeterminate sources. This relates to the redshift range of the sample: ULIRGs are rare sources but with a higher presence in the high-z Universe. Since the radio sample is biased to lower redshift by the survey sensitivity, it is to be expected that there should be larger numbers of luminous galaxies in the entire sample of the wider GMRT survey. This is corroborated with average redshift; for the higher luminosity sources are on average at higher redshift: $z_{ULIRGs}$ = 0.88, $z_{LIRGs}$ = 0.66, whereas for the lower luminosity sample is $z$ = 0.26.

Following from the FIR-radio correlation, the radio-SFR conversion allows us to convert the IR luminosity into radio luminosity (\citealt{condon02}).
\begin{equation}
L_{\nu}\left(1.4~{\rm GHz}\right) = 4\pi D _{L}^2\left (1 + z\right)^{-\left ( 1 + \alpha\right)}\left ( \frac{1.4~{\rm GHz}}{0.61~GHz} \right )^\alpha F_\nu (610~{\rm MHz})
 \label{radio_fit_eqn}
\end{equation}
where $L_{\nu}\left(1.4~{\rm GHz}\right)$ and $L_{\nu}\left(610~{\rm MHz}\right)$ are the 1.4 GHz and 610 MHz luminosities, $D _{L}^2$ is the luminosity distance, $z$ is the spectroscopic redshift and the spectral index $\alpha$ = -0.8. 

Following the same procedure as for the FIR-radio correlation, a source with F$_{1.4~\rm {GHz}}$ = 1 mJy at z~=~0.3 would have a radio luminosity of 2.6 $\times$ 10$^{23}$ W Hz$^{-1}$. According to the conversion used by \cite{serjeant2002}, this luminosity would correspond to an SFR $\sim$ 300~\Msun~yr$^{-1}$ which is well within the range for starburst galaxies.

The presence of a central AGN (AGN presence is defined from the contributions of AGNs to the overall spectral energy distribution (SED) of galaxies, where the fractional contribution of the AGN over the wavelength range 5-40 $\mu$m, compared to the total infrared luminosity over the range 8-1000 $\mu$m.) in a dusty galaxy can have a significant effect on the mid-infrared spectrum due to the extreme heating effect of the AGN emission (accretion disk). The dust torus heated by the central AGN emits strongly over the 5-30$\mu$m range with the dust being heated to its sublimation temperature, resulting in the destruction of small grains (including PAHs) and a featureless shallow power law spectrum as discussed by \cite{laurent2000}. Analysis of these mid-infrared colours can be used to distinguish AGN from star-forming galaxies as discussed by \citet{brand06}.

It is also known that the presence of an AGN can regulate the properties of the galaxy \citep{bower06}, and it is noted that this radio sample contains some evidence of AGN activity in 56$\%$ of the sources to fit with CIGALE, but only 4$\%$ are AGN dominated, contributing to the integrated emission from 20$\%$ to 38$\%$. The average AGN fraction for the total sample is 7$\%$ and the explanation for this unexpectedly low AGN contribution is the low radio flux. Generally in the sample, the AGN fraction shows a trend with luminosity with higher AGN fractions corresponding to higher redshift and luminosity. To evaluate this possible correlation, the LIRGs sample was selected (to mitigate against Malmquist bias) and four redshift bins were produced, showing that there is a weak correlation between the AGN fraction and the redshift. (see Figure \ref{LFIR}). However, the AGN fraction is not dominant in the total sample and the percentages are lower than 10$\%$. It is necessary to evaluate the sample by luminosity: both ULIRGs and sources with $L_{IR}$ $<$~10$^{11}$\Lsun ~show that 50$\%$~have an AGN present, whereas LIRGs have the highest AGN presence (of the sample (59$\%$) and galaxies at lower luminosities show lower AGN presence (42$\%$). The AGN contribution in each subsample is six, five and two per cent for ULIRGs, LIRGS, and lower luminosity sources respectively. Therefore in this sample, the AGN presence does not seem to be convincingly related to the luminosity, however, at higher luminosity, for those sources that host an AGN, the contribution of the AGN is two to three times more important than in the less luminous galaxies.

There is currently no definitive agreement on whether AGNs trigger or quench star formation, while some studies as \citet{barger15} suggest that star formation can be suppressed by the AGN, other authors have inferred the opposite (e.g. \citealt{juneau13}). However, it is accepted that both phenomena should be related (e.g. $\sigma$ -- M$_{SMBH}$, the empirical correlation between the stellar velocity dispersion of a galaxy bulge and the mass of the supermassive black hole from \citet{F2000} and \citet{G2000}) and there is a similarity between the growth of the AGN and the star formation history. The star formation history (SFH) chosen is the delayed module in CIGALE which assumes a star formation rate that decreases exponentially over cosmic time. Since the SFR resulting from CIGALE depends on the models introduced on the star formation history (SFH), a common SFH representative of moderate redshift galaxies with no starburst population was assumed. The star formation rate of the radio sources is relatively low, SFR = 24 $\pm$ 4~[\Msun~/~yr] on average, probably due to the lower redshift sample. SFR is proportional to the far-IR luminosity, therefore ULIRGs have the highest value of the sample with SFR$_{ULIRGs}$ = 143 $\pm$ 4~[\Msun~/~yr], more than 5 times that of the LIRGs and 37 times more than the rest of sources.

\begin{figure}
 \centering
 \includegraphics[angle=-90,width=1\linewidth]{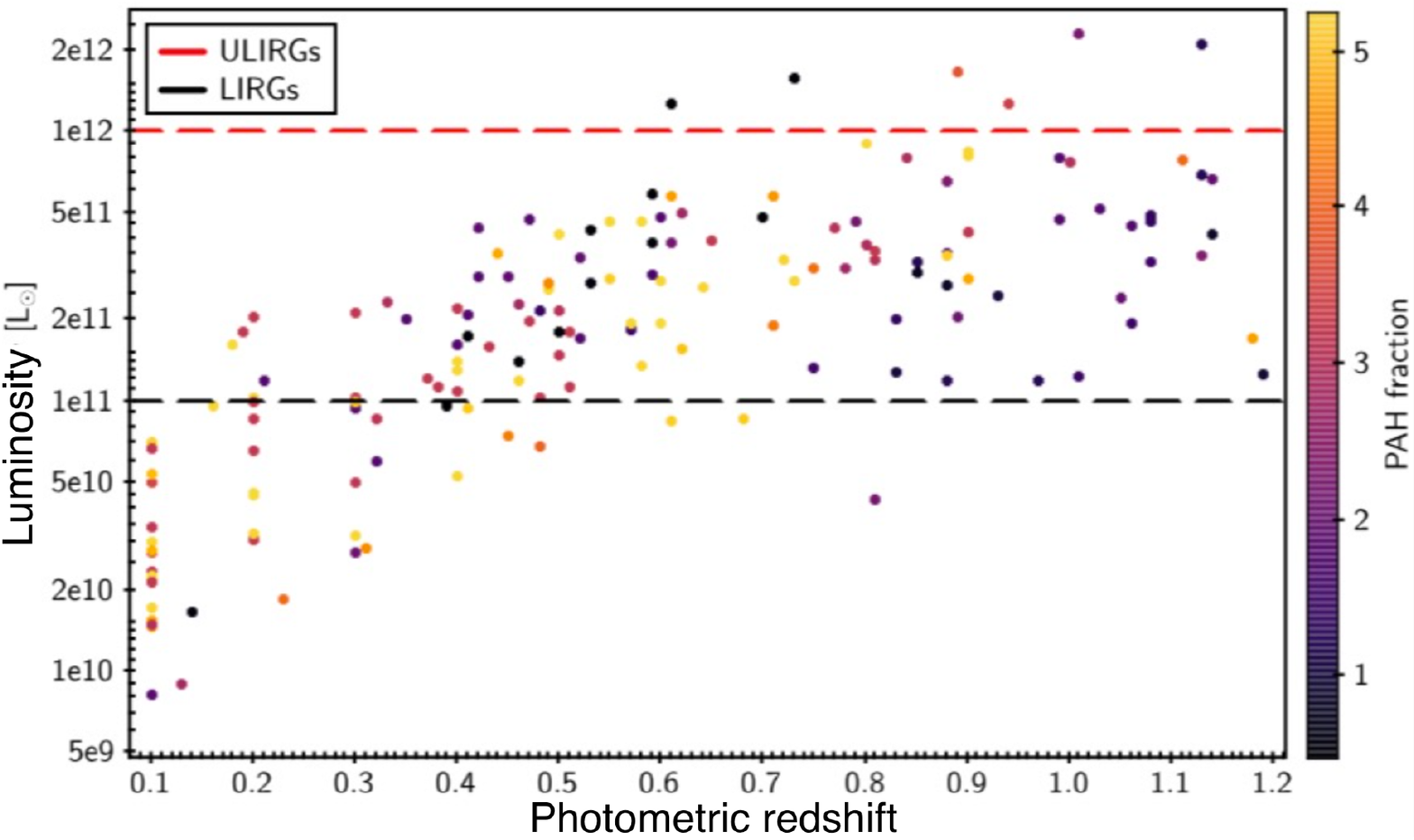}\\
 \caption{This plot shows the infrared luminosity against photometric redshift as in the previous Figure but with the PAHs fraction in the third axis. There is no clear correlation between PAHs and infrared luminosity nor with redshift. However, on average, the ULIRGs - sources above the dashed red line - appear to have only about half of the PAHs as galaxies with L$_{IR}$ $<$10$^{11}$ \Lsun~(sources below the dashed black line) The photometric errors are between 0.005 $<$ $z{_{phot}}$ $<$ 0.06, whereas the uncertainty in the IR luminosity corresponds to an average of 11$\%$.}
\label{LPAH}
\end{figure}

The role of the PAHs in star formation through the comparison between luminosity and PAH mass fraction has been discussed by \citet{dl07}, and is a parameter $q_{PAH}$ in the CIGALE modelling (see Table \ref{CigaleSEDinputs}), where a normalisation or scaling process takes into account the IR luminosity of the galaxy, ensuring that the derived PAH properties reflect the underlying physical conditions of the galaxy, considering variations in luminosity and the associated dust emission. was examined (see Figure \ref{LPAH}).

There is no clear correlation between the luminosity of the galaxy and the presence of PAHs. However, on average ULIRGs appear to have only about half of the PAHs as galaxies with luminosity below 10$^{11}$ \Lsun, which together with the higher redshift of this subsample (see Table \ref{LIRparameters}), indicates that part of the high luminosity could come from AGN activity instead of star formation.

\begin{table}
\centering
\begin{tabular}{|c|c|c|c|}
\hline
 & \textbf{ULIRGs} & \textbf{LIRGs} & $ L < 10^{11}L_{\odot}$ \\\hline
Percentage of sources & 4\% & 66\% & 30\% \\
AGN presence & 50\% & 59\% & 42\% \\
Average of AGN fraction & 6\% & 5\% & 2\% \\
SFR [M$_{\odot}$/yr] & 143 & 26 & 3.8 \\
$M_{*}$ [M$_{\odot}$] & $4.8 \cdot 10^{11}$ & $1.43 \cdot 10^{11}$ & $5.19 \cdot 10^{10}$ \\
$M_{gas}$ [M$_{\odot}$] & $3.5 \cdot 10^{11}$ & $1.1 \cdot 10^{11}$ & $4.14 \cdot 10^{10}$ \\
Photometric redshift & 0.88 & 0.66 & 0.26 \\
\hline
\end{tabular}
\caption{This table shows the main parameters from the IR luminosity classification: ULIRGs, LIRGs and lower infrared luminosity galaxies, showing that the dominant population is high infrared luminosity (LIRGS). More than half of the sources need some AGN contribution in the fit, however, they present low AGN fractions, which means they are not AGN-dominated. The SFR is significantly higher in higher luminosity sources as was expected. The stellar mass and gas mass is one order of magnitude higher in high infrared luminous galaxies, which may indicate that the higher luminosity and SFR are a consequence of high stellar emission - although higher quality data are needed to be sure of this conclusion.}
\label{LIRparameters}
\end{table}

The data can be used to examine the use of infrared colour-colour diagrams to classify AGN in the NEP. using colour-colour diagrams to segregate populations using ${\it AKARI}$ , Spitzer/IRAC \citep{do12} and/or using ${\it WISE}$  data (e.g. \citealt{la12}; \citealt{Mingo16}; \citealt{solarz15}). Previous studies referenced in the ancillary data table have shown the presence of two main extragalactic populations: local star-forming galaxies at z $\sim$ 0.25, and luminous infrared galaxies (LIRGs) at z $\sim$ 0.9.

\begin{figure*}
 \centering
%%\hspace*{-1 cm}
 %%\includegraphics[angle=0,width=1.2\linewidth]{GMRT_N2_N4_S7_S11.pdf}\\
 \includegraphics[angle=-90,width=1\linewidth]{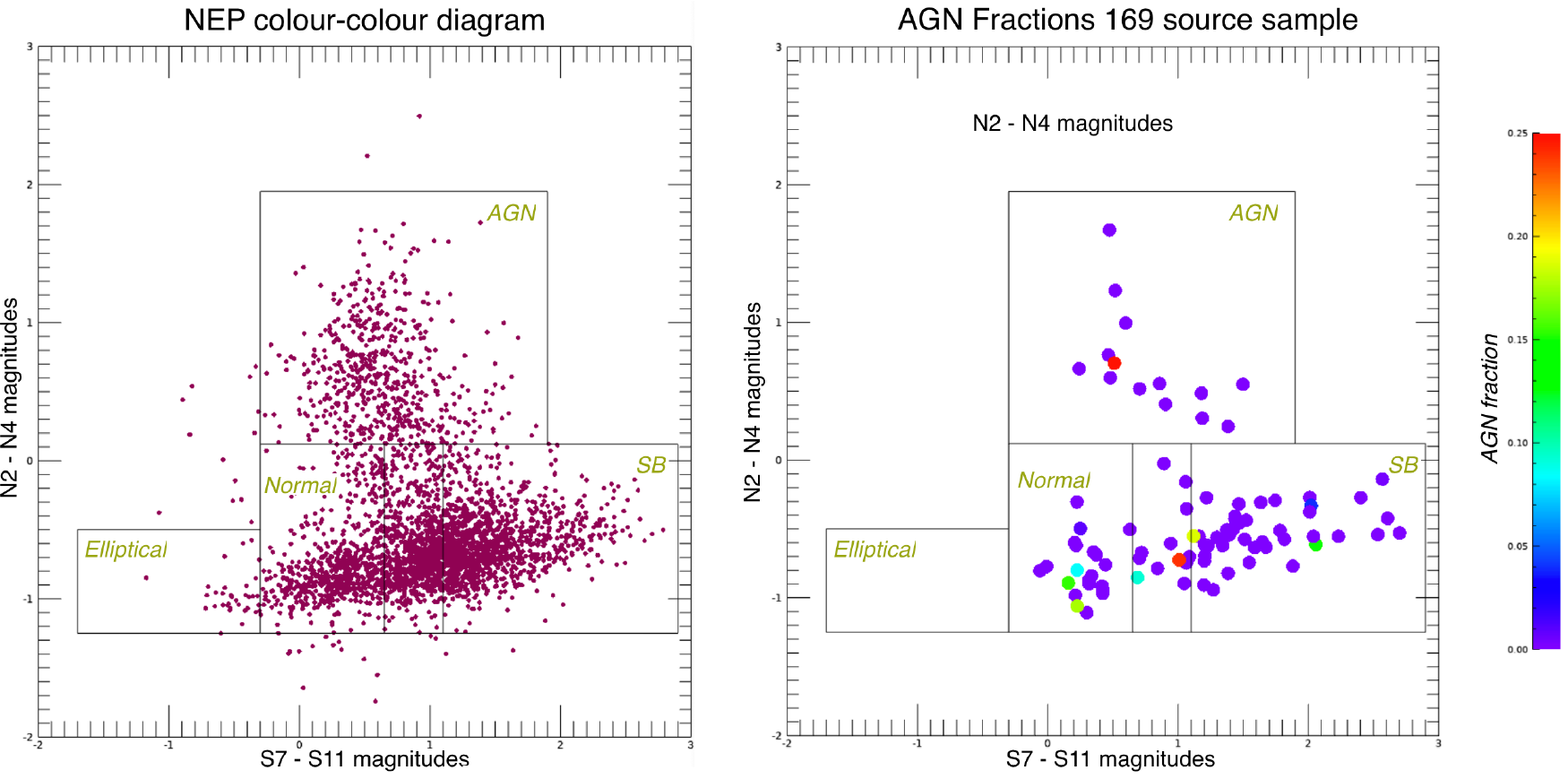}\\
 \caption{The [left] plot shows the complete ${\it AKARI}$  NEP colour-colour diagram from all infrared sources: plotted as N2- N4 colour against S7-S11, which allows separation of the AGNs from normal galaxies. Following the conditions (N4 - N2) $>$ 0.1 and 1.5 $<$ $S7 - S11$ $<$ -0.3 about 90$\%$ of the sources are likely to be AGN (69$\%$ of which have been spectroscopically classified). The [right] plot shows the locations of sources which we have optically identified with the centroid of the GMRT sources, with colour coding to show the estimated AGN fraction based on CIGALE modelling. In those cases where the GMRT sources are grouped in PyBDSF, we use the assumed central source. The relative populations between the different kinds of galaxies are broadly consistent between the ${\it AKARI}$  and radio-selected samples, except that the radio sample has, as expected, a higher proportion of AGN (27$\%$ for the radio sample compared with 17$\%$ in the ${\it AKARI}$  sample).}
\label{AKARIbestcc}
\end{figure*}

The advantage of using ${\it AKARI}$  data is that the satellite was equipped with multiple filters, improving the previous result of nearly 80$\%$ effectiveness in classifying AGN. The colour (S7 - S11) used earlier following the \cite{shim13} spectroscopic work also showed good results. Therefore, different colours were checked on the x-axis of the colour-colour diagram finding that shorter wavelengths better identify the AGN in a particular colour-colour area. This condition set the power-law condition at longer wavelengths, furthermore, it confines the spectroscopic AGN in the same region. Applying the condition (N2 - N4) $>$ 0.1 - with 0 $<$ S7 - S11 $<$ 1.5, 69$\%$ of the sources are spectroscopically classified as AGN. The use of two different colours with close wavelengths (4.4~$\mu$m - 2.4~$\mu$m and 10.9~$\mu$m - 7.3~$\mu$m) allows the identification of two different parts of the mid-IR spectra. This is a clear advantage concerning ${\it WISE}$  data. since with only four bands is not possible to select two colours with such close bands. The two colours allow detection of the prominent drop in galaxy emission and the power-law spectrum in two different parts of the mid-IR, allowing the location of the AGNs to be found in a narrow region in the colour-colour space (see Figure \ref{AKARIbestcc}), which makes the selection more effective.

\subsection{Summary of the 169 source sample}
It is useful to summarise this small, non-statistical selected sample of mostly sub-mJy sources selected from the GMRT survey as having substantial data at other wavelengths to allow modelling of the source characteristics.  The photometric redshifts vary between $z$ $\sim$ 0.1 - 1.25, which the highest concentration around $z$ = 0.1.in the 169 source sample luminous infrared galaxies (LIRGs) represent represent 66$\%$ of the sample, ultra-luminous infrared galaxies (ULIRGs) 4$\%$ and sources with L$_{IR}$~$<$~10$^{11}$ \Lsun~ 30$\%$. In total 56$\%$ of sources require some AGN presence, although only seven sources are AGN-dominated. The spectral index of the sources showed typical synchrotron emission characteristics with some sources exhibiting spectral turnovers at lower frequencies, which possibly indicates variability or alternative explanations like free-free absorption or non-standard electron energy distributions.

\section{Radio-FIR properties}
\label{radiofircorrelation}

The Radio-FIR correlation in the NEP field has been studied from a catalogue obtained by cross-matching the ${\it AKARI}$  data \citep{mu13} with FIR data from ${\it HERSCHEL}$  \citep{pearson18} and the present GMRT radio data. These catalogues were cross-matched using an optical catalogue as a reference and half of the synthesised beam size as a search radius. A 9\arcsec~search radius was used for the SPIRE catalogue using the 250~$\mu$m position and 4\arcsec~for the radio position. Furthermore, the photometric redshift catalogue of \cite{Oi2014} was included, to calculate the radio luminosity.

To examine the Radio-FIR correlation, the rest-frame luminosity density at 1.4 GHz based on the GMRT 610 MHz flux density is calculated for the 169 source sample using Equation \ref{radio_fit_eqn}: The total IR luminosity (TIR) was calculated from the CIGALE best-fit model from rest-frame 8-1000 $\mu$m, without any correction for the radio luminosity. Quenched galaxies (i.e. those having significantly reduced or ceased star formation activity less than about 1 \Lsun yr$^{-1}$ or a specific star formation rate less than 10$^{-11}$ yr$^{-1}$), AGN (from our CIGALE analysis), and outliers with reduced $\chi^2$ $<$ 5 were excluded from the linear fit.
\begin{figure}
 \centering
 \includegraphics[angle=0,width=1.1\linewidth]{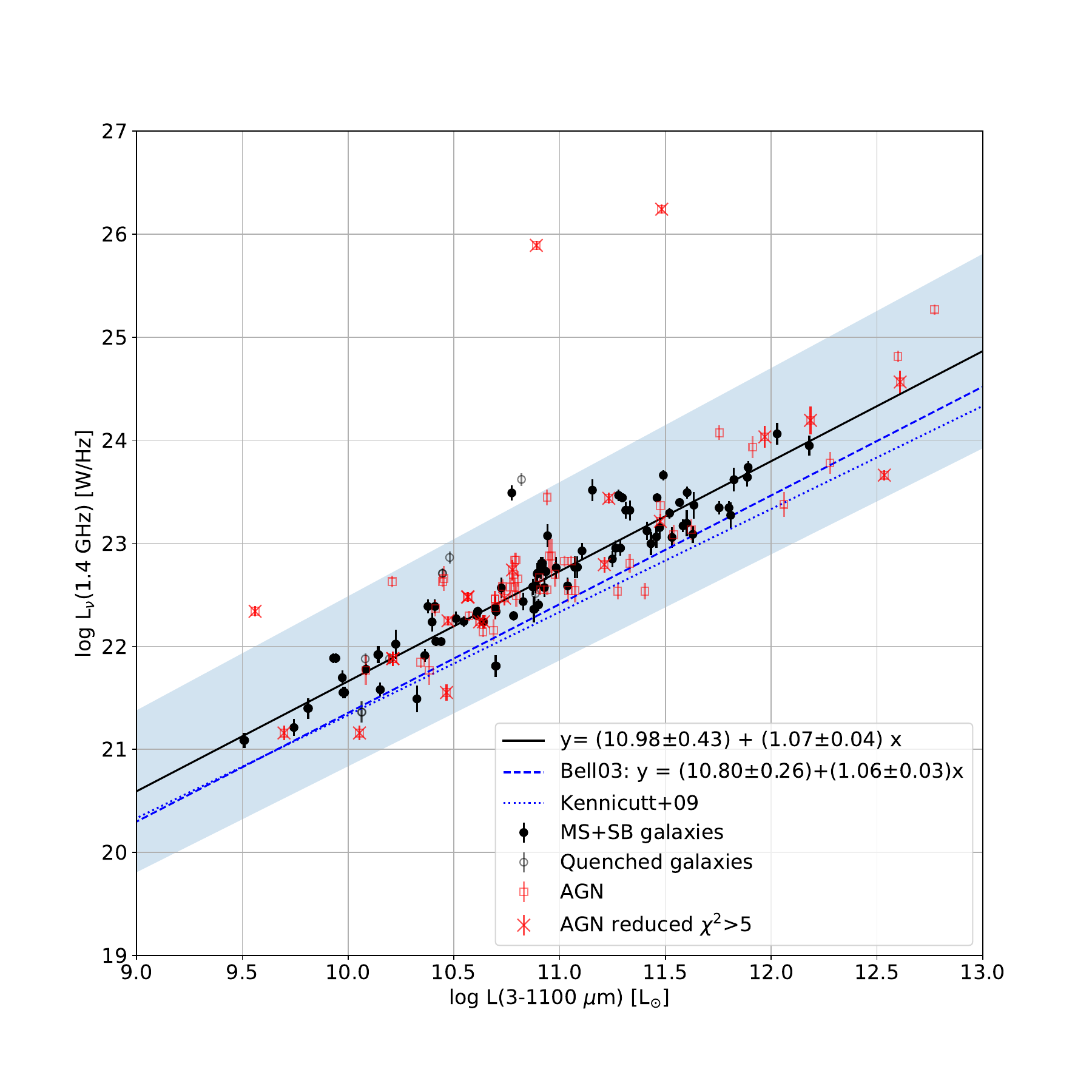}\\
 \caption{Radio-FIR correlation between the GMRT 610 MHz luminosity density, scaled to the more commonly reported 1.4 GHz, and the total infrared luminosity, along with error estimates as described in the text. In the caption 'MS' stands for main-sequence and 'SB' stands for starburst. This graph, also published in \citealt{Kim23}, uses the 610 MHz fluxes to estimate the 1.4 GHz luminosity density shown on the vertical axis as calculated in Equation \ref{radio_fit_eqn}.}
\label{radiofircorrelation}
\end{figure}

The fit to the main-sequence and starburst sample shown in Figure \ref{radiofircorrelation} is linear and consistent with the local relation.

\begin{figure}
 \centering

 \includegraphics[angle=-90,width=1.0\linewidth]{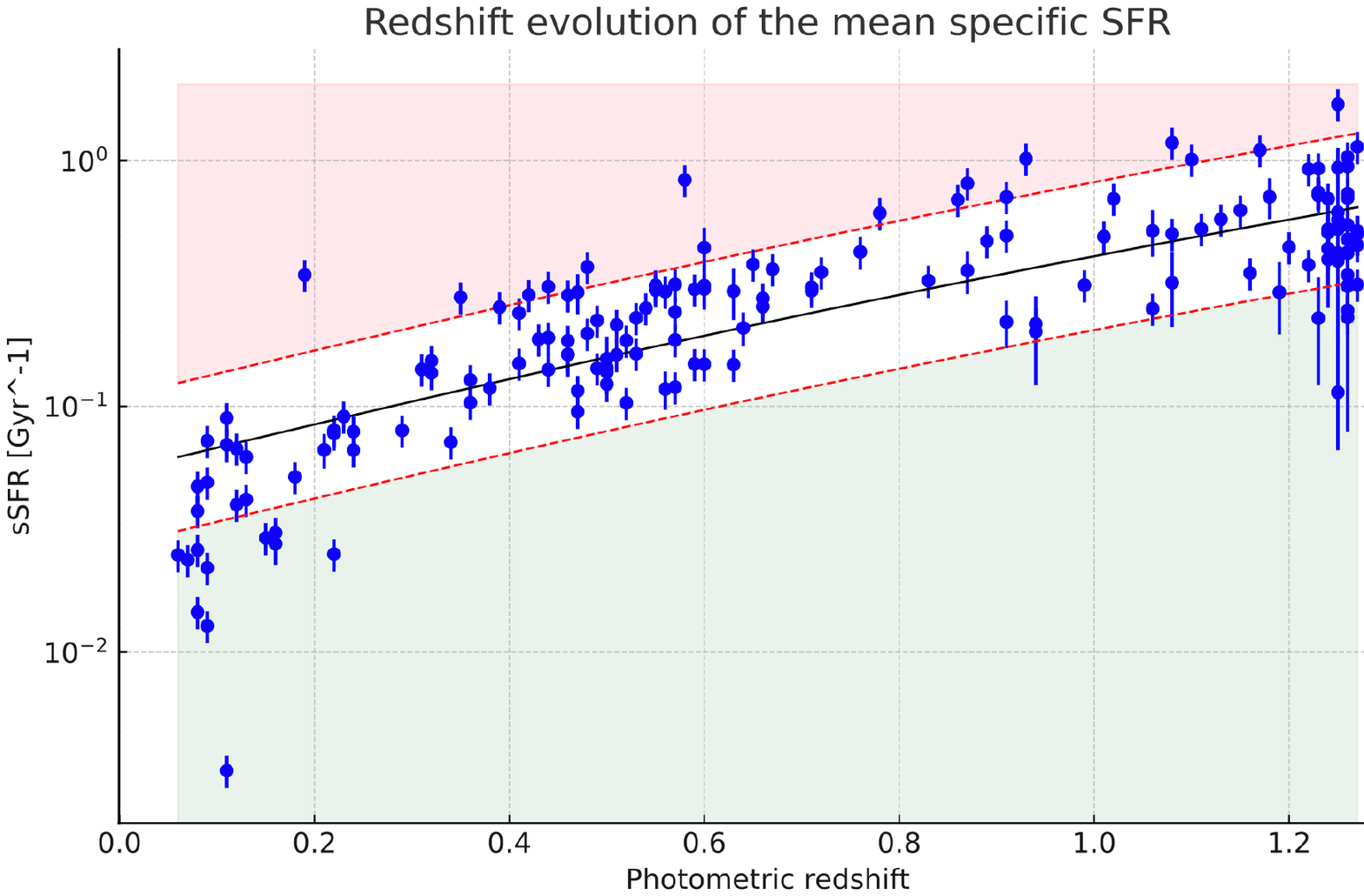}\\
 \caption{This shows the redshift evolution of s${\it SFR}$ for galaxies in the 169 source sample. The central solid line shows the expected redshifted evolution of ${\it SFR}$ from \citet{El11}, whilst the upper and lower dotted lines show values a factor of 2 times higher and lower than this respectively. Galaxies above the upper dotted line, in the red-shaded section, are likely starburst galaxies, whilst those below the lower dotted line in the green-shaded section have lower s${\it SFR}$  value. The error bars are shown to be representative of the likely uncertainties.}
\label{ssfr}
\end{figure}

Following \cite{El11} from studies in the GOODS field, we assume that the slope of the ${\it SFR}$ – M$_{\ast}$ relation is equal to 1 at all redshifts, hence that the specific SFR, s${\it SFR}$ (= SFR/M$_{\ast}$), is independent of stellar mass at fixed redshift. This allows us to define a 'Main Sequence' for normal galaxies. We show a plot in Figure \ref{ssfr} of the s${\it SFR}$ for sources from the 169 source sample using the estimated photometric redshifts from the CIGALE modelling.

To first order the 169 source sample follows this expected 'Galaxy Main Sequence' relationship with a few galaxies showing higher rates of star formation or nuclear activity, otherwise, there is nothing remarkable about this small sample of the larger radio survey detections that suggests a strong bias.

\section{Associations with other sources}
\subsection{Quasars, Seyfert Galaxies and X-ray sources}
Cross-matching the GMRT data with the Half Million Quasars (HMQ) Catalogue (\citealt{flesch2015}) led to 13 cross-matches (3 with spectroscopic, and 4 with photometric redshifts). Of these nine were found within 1.2\arcsec~of a GMRT radio component, three within 1.7\arcsec,~and one within 2.8\arcsec, and redshifts extending out to 2.1. In the related associations, with the MORX catalogue (\citealt{flesch2016}) that associates optical objects with radio and X-ray sources, we find 45 cross matches, with 16 ${\it CHANDRA}$ , 10 XMM, 6 SWIFT and 8 other X-ray detections, or which there are 23 having previously known VLA associations.

One galaxy cluster X-ray source is located within the NEP area, for which five spectroscopic redshifts show this cluster to be at a redshift $z$~=~0.6909. Two of the cluster galaxies (cfh 2 at RA$_{2000}$ = 17$^{\rm h}$ 57$^{\rm m}$ 46$^{\rm s}$.3, Dec$_{2000}$ = +66$^{\circ}$ 30' 26", $z$ = 0.7006; and cfh 8 at RA$_{2000}$ = 17$^{\rm h}$ 57$^{\rm m}$ 29$^{\rm s}$.9, Dec$_{2000}$ = +66$^{\circ}$ 32' 29", $z$ = 0.6860) have been identified as AGN2 galaxies with their optical spectra showing narrow forbidden [O III], [O II] and permitted H${\beta}$ and H${\gamma}$ emission lines. In total thirteen of the thirty-seven GMRT/${\it CHANDRA}$  sources cross-match with AGNs detected in all four ${\it AKARI}$  N-filters, with nine having longer wavelength ${\it AKARI}$  detections. 

We also confirm the detection reported by \cite{branchesi06} at 1.4 GHz of the redshift 0.691 cluster/X-ray source NEP200, which has a GMRT flux density of 243 $\pm$ 59 $\mu$Jy. Compared to the reported \cite{branchesi06} 1.4 GHz flux of 342 $\pm$ 30 $\mu$Jy, this suggests that either the source has varied in the $\sim$ 5 years between the VLA and GMRT observations, or that the spectral index is unusually inverted.

One notable exception was the lack of any detectable radio emission from the nearby X-ray source/AGN J175714.5+663112.9 which is at a spectroscopic redshift of 0.4597 (\citealt{shim13}).
\subsection{Sub-mm galaxy associations}
\label{Submm}

Observations have recently been reported of two 850 $\mu$m sub-mm wavelength surveys of the NEP region using the JCMT: S2CLS from \cite{geach17} containing 196 detections at $>$ 4$\sigma$ with $<$ 6$\%$ false detection probability in a 0.6 square degree area, and NEPSC2 from \citep{shim20}  with 549 sources at $>$ 4$\sigma$ with $<$ 10$\%$ false detection in a two square degree area and 342 sources at $>$ 4.5$\sigma$ with a $<$ 3$\%$ false detection probability. By combining the two surveys, the number of unique 850-\mum sources is 647, for which multi-wavelength counterparts are identified using ancillary data including deep optical, NIR photometry (\citealt{Oi2021}) and high-resolution VLA 1.5 GHz data \citep{shim22}.

\begin{figure} 
 \centering
 \includegraphics[angle=0,width=1\linewidth]{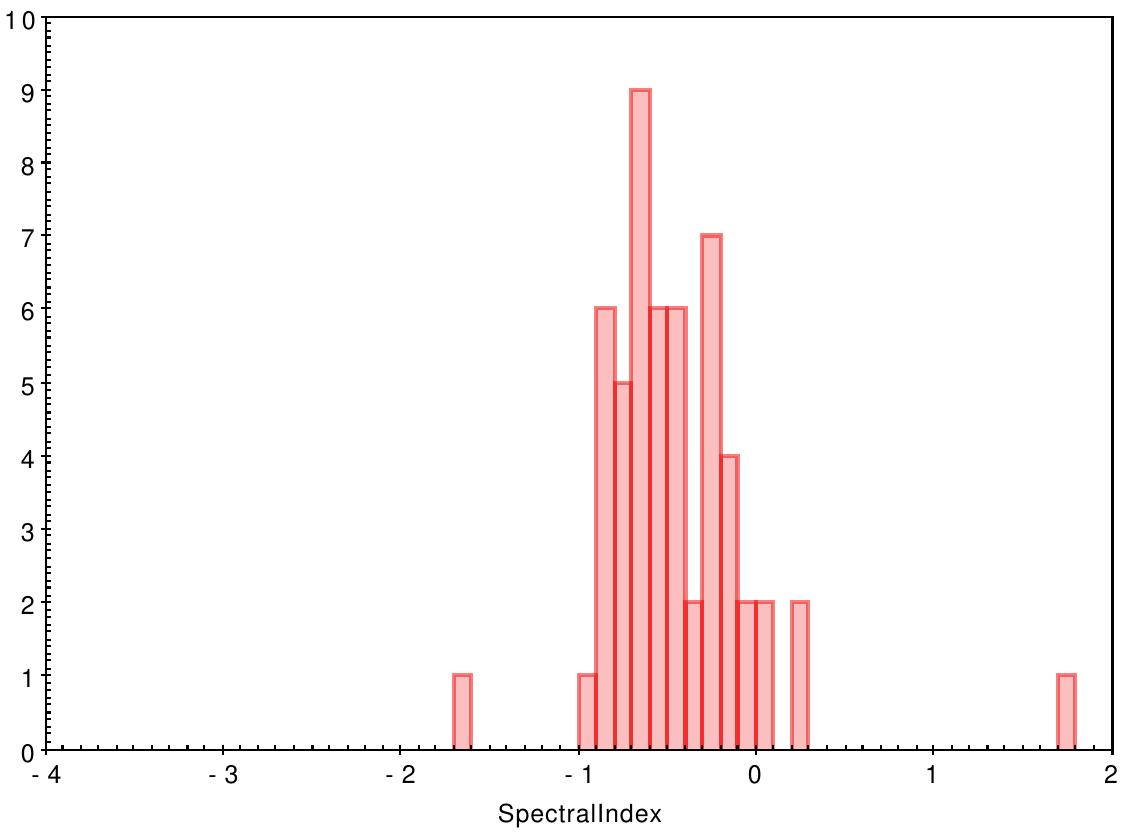}\\
 \caption{This shows the GMRT 610 MHz - LOFAR 135 MHz spectral index for the NEPSC2 sub-mm associations, with the vertical axis showing the number of detections in a bin. For presentation purposes, the data point for the Catseye Nebula has been removed from this Figure.}
\label{lofar_gmrt_NEPSC2_spectral_index}
\end{figure}

About 80$\%$ of the GMRT sources are located in the area covered by JCMT observations. With a matching radius of 1 arcsec, 61 cross-matches were found. Among these, 34 objects are radio-identified sub-mm galaxies detected in 1.5 GHz. The mean photometric redshift of 61 matched sources is $<$z$>$ = 1.98 $\pm$ 0.14 and the mean IR luminosity is $\langle \mathrm{log}(L_{IR}/\mathrm{L}_\odot) \ge 12.43 \pm 0.12$. Sub-mm associated 610 MHz sources are the most IR-luminous population among the entire 610 MHz sources, most of which are located toward higher redshifts. The mean radio spectral index of the sub-mm associations is -0.52$\pm$0.09 and -0.45$\pm$0.10 for 610-135 MHz (Figure \ref{lofar_gmrt_NEPSC2_spectral_index}) and 1400-610 MHz respectively, close to that of the starburst population. 

The brightest extragalactic GMRT source that is cross-matched to the sub-mm source catalogue is the GMRT source J175335.23+663546.31 (S$_{850}$ = 5.1 mJy), with an integrated flux density of 218 mJy. This is a known radio source (NVSS J175335+663546) with S$_{1.5GHz}$ = 186 mJy and is associated with an R = 21.5 magnitude optical source having a relatively red optical colour. The photometric redshift is around 0.6 (\citealt{Oi2014, ho2021}), and the source is also detected at X-ray wavelengths (\citealt{Krumpe2015}). Therefore, this source is considered to be an early-type galaxy hosting a radio-loud AGN.

.

\section{Conclusions}
\begin{enumerate}
\item The GMRT 610 MHz survey data reported in this study had a 3.6\arcsec $\times$ 4.1\arcsec~synthesised beam and achieved an average 5$\sigma$ sensitivity of the survey averaged over the whole area after primary beam correction was 28~$\mu$Jy beam$^{-1}$. The more sensitive central area has a sensitivity reaching as low as 19~$\mu$Jy beam$^{-1}$, resulting in a final catalogue of 1675 radio components with integrated fluxes above the $\sim$100~$\mu$Jy level, with 339 of these grouped into multi-component sources, with an additional 284 subsequently believed to be part of double radio sources.
\item The GMRT data are compared with new 135 MHz LOFAR High Band Array data to estimate the spectral index of the brighter sources, resulting in 678 cross-matches, of which 255 radio components were also detected with the WSRT. Examining the double spectral indices from the WSRT-LOFAR and GMRT-LOFAR data, most sources lie, as expected, in the range of steep spectral shapes characteristic of synchrotron emission. There is however evidence for a small number of inverted spectra, the most obvious of which is associated with a previously identified QSO. This may indicate variability between the different survey epochs, or evidence for an alternative explanation such as free-free absorption, or non-standard energy distribution of relativistic electrons which differs from a power law, such as a Maxwellian distribution.
\item The classification of the radio sample by infrared luminosity gives a high number of luminous sources: LIRGs represent 66$\%$ of the sample, whereas the ULIRGs are only 4$\%$ and sources with L$_{IR}$~$<$~10$^{11}$ \Lsun~make up the remaining 30$\%$.
\item The radio sample contains 56$\%$ of sources that require some AGN presence at some level for the SED fitting, although the PAH mass fraction is found not to be a significant indicator of infrared luminosity or redshift in the small 169 source sample.
\item The AGN presence was evaluated in the LIRGs population, showing an increasing trend of the AGN fraction with redshift, that if continuous it would peak at z $\sim$ 2, which is the peak of star formation and supermassive black hole growth. The PAH concentration is not significant, however, on average ULIRGs appear to have only about half of the PAH strength than galaxies with lower IR luminosities, finding an expected PAH deficit in the most luminous sources which tend to be at higher redshift in the sample
\item The gas mass (and the stellar mass) are one order of magnitude larger in higher luminosity galaxies (both LIRGs and ULIRGs) than in lower luminous galaxies. These higher masses indicate higher star formation rates, tentatively suggesting that the high luminosities could proceed from the star formation instead of an AGN presence.
\item For the LIRGs there is an evolution of the AGN fraction with redshift indicating that at high IR luminosities (therefore at high SFR) the AGN presence is higher at higher redshifts. Together with the PAH deficit and the AGN fraction, these results indicate that part of the luminosity could be contributed by AGN activity instead of star formation.
\item Cross-matching data from the Giant Metrewave Radio Telescope (GMRT) with the Half Million Quasars (HMQ) Catalogue, revealed 13 cross-matches. The brightest identified GMRT association is the Seyfert 2 galaxy NGC 6552, while another notable object is the QSO source 7C 1804+6625. Cross-matching with the MORX catalogue finds 45 cross-matches, including various X-ray detections. Notably a bright galaxy cluster X-ray source within the NEP area was found that has a redshift of z=0.6909, and two of the cluster galaxies were identified as AGN2 galaxies, displaying specific emission lines in their optical spectra. Thirteen GMRT/${\it CHANDRA}$  sources cross-matched with AGNs detected in all four ${\it AKARI}$  N-filters. The lack of detectable radio emission from the X-ray source/AGN J175714.5+663112.9 was also notable. Amongst the sub-mm galaxy associations, 61 cross-matches were found, of which 34 objects are radio-identified sub-mm galaxies detected in 1.5 GHz. Finally, the brightest extragalactic GMRT source cross-matched to the sub-mm source catalogue is GMRT source J175335.23+663546.31. It is considered to be an early-type galaxy hosting a radio-loud AGN.
\end{enumerate}

\section{Acknowledgements}

GJW  gratefully acknowledges the support of an Emeritus Fellowship from The Leverhulme Trust. DP acknowledges the post-doctoral fellowship of the S. N. Bose National Centre for Basic Sciences, Kolkata, India, funded by the Department of Science and Technology (DST), India. KM is grateful for support from the Polish National Science Centre via grant UMO-2018/30/E/ST9/00082. TN was supported by JSPS KAKENHI Grant Number 21H04496 and 23H05441. We thank the staff of the GMRT that made these observations possible. The GMRT facility is run by the National Centre for Radio Astrophysics of the Tata Institute of Fundamental Research.

%%%%%%%%%%%%%%%%%%%%%%%%%%%%%%%%%%%%%%%%%%%%%%%%%%
\section*{Data Availability}

The data underlying this article are available in the article and in its online supplementary material.

\label{lastpage}

\end{document}